\documentclass[
aps,
pra,
reprint,
superscriptaddress
]{revtex4-2}

\pdfoutput=1

\usepackage{amsmath}
\usepackage{amsfonts}
\usepackage{amssymb}
\usepackage{tikz}
\usetikzlibrary{quantikz}
\usetikzlibrary{shapes.geometric}
\usepackage{caption}
\usepackage[percent]{overpic}
\usepackage{blindtext}
\usepackage[colorlinks=true,
linktocpage=true
]{hyperref}
\usepackage{url}

\begin{document}

\newcommand{\AM}[1]{\textcolor{blue}{#1}}

\title{Optimizing one-axis twists for variational Bayesian quantum metrology}

\author{Tyler G. Thurtell}
\email{tthurtell@unm.edu}
\affiliation{Department of Physics and Astronomy, Center for Quantum Information and Control, University of New Mexico, Albuquerque, New Mexico 87106, USA}

\author{Akimasa Miyake}
\email{amiyake@unm.edu}
\affiliation{Department of Physics and Astronomy, Center for Quantum Information and Control, University of New Mexico, Albuquerque, New Mexico 87106, USA}

\date{March 15th, 2024}

\begin{abstract}
Quantum metrology and sensing seek advantage in estimating an unknown parameter of some quantum state or channel, using entanglement such as spin squeezing produced by one-axis twists or other quantum resources. In particular, qubit phase estimation, or rotation sensing, appears as a ubiquitous problem with applications to electric field sensing, magnetometry, atomic clocks, and gyroscopes. By adopting the Bayesian formalism to the phase estimation problem to account for limited initial knowledge about the value of the phase, we formulate variational metrology and treat the state preparation (or encoding) and measurement (or decoding) procedures as parameterized quantum circuits. It is important to understand how effective various parametrized protocols are as well as how robust they are to the effects of complex noise such as spatially correlated noise. First, we propose a new family of parametrized encoding and decoding protocols called arbitrary-axis twist ansatzes, and show that it can lead to a substantial reduction in the number of one-axis twists needed to achieve a target estimation error. Furthermore, we demonstrate that the estimation error associated with these strategies decreases with system size in a faster manner than classical (or no-twists) protocols, even in the less-explored regimes where the prior information is limited. Last, using a polynomial-size tensor network algorithm, we numerically analyze practical variational metrology beyond the symmetric subspace of a collective spin, and find that quantum advantage persists for the arbitrary-axis twist ansatzes with a few one-axis twists and smaller total twisting angles for practically relevant noise levels.
\end{abstract}

\maketitle

\section{Introduction}
Sensors that utilize quantum mechanical effects to increase sensing precision and accuracy are expected to be among the first quantum information technologies to achieve broad application \cite{Degen_2017}. The study of the abilities of such sensors is referred to as quantum sensing or quantum metrology. When the system being used as a sensor is a collection of qubits or spin-1/2 particles a quintessential problem is qubit phase estimation, or rotation sensing~\cite{Degen_2017, Kitching2011}. Many applications can be understood as instances of this problem including magnetometry with neutral atoms~\cite{Budker2007}, atomic clock stabilization with trapped ions~\cite{Diddams2001} or with neutral atoms~\cite{Clairon1995, Shirley2001}, nuclear magnetic resonance gyroscopes~\cite{Donley2010}, Rydberg atom electric field sensors~\cite{Mohapatra2008}, and many other examples. This paradigm may even be relevant to quantum algorithms~\cite{Pezze2021}.

It has been known for a long time that using entangled many-body states can in principle lead to sensing precision that would not have been possible with an unentangled state of the same system \cite{Bollinger_1996, Giovannetti_2004, Giovannetti_2006, Giovannetti_2011, Toth_2014, Degen_2017, Pezze_2018}. Typically, the available advantage is quadratic in system size, leading to the famous Heisenberg limit \cite{Giovannetti_2006}. A useful type of entanglement for rotation sensing with spin-1/2 particles is spin-squeezing which can be generated, for example, via one-axis twisting type interactions~\cite{Wineland1992, Kitagawa_1993}. These interactions have been engineered in Bose-Einstein condensates~\cite{Henkel_2010}, cavity QED setups~\cite{Braverman_2019}, trapped ions~\cite{Pogorelov2021}, and approximately in Rydberg atom arrays~\cite{Pupillo_2010, Gil_2014}. When one-axis twisting is used in only the state preparation phase the available advantage scales with system size to the power of 5/3~\cite{Kitagawa_1993}. If one-axis twists are used in both the state preparation and measurement phases, as in twist-untwist protocols~\cite{Leibfried_2004, Leibfried_2005, Davis_2016, Frowis_2016, Macri_2016, Nolan_2017, Colombo_2022, Li_2022}, the available advantage scales quadratically. These protocols have been realized experimentally for example in~\cite{Colombo_2022}.

However, the situation is further complicated when prior knowledge about the value of the parameter to be estimated must be accounted for \cite{Buzek_1999, Berry_2000, Macieszczak_2014, Kaubruegger_2021, Friis_2017, Wolk_2020, Jarzyna_2015, Gorecki_2020, Holevo_1980, Personick_1971, Martinez-Vargas_2017, Tsang_2020, Bonato2016, Rosenband2013, Leroux2017}. In addition, the extreme sensitivity of the required states to noise has so far prevented entanglement enhanced sensors from becoming a practical reality \cite{Huelga_1997, Escher_2011, DemkowiczDobrzaski_2012, Jarzyna_2013, Koodyski_2013, Tsang_2013, Yamamoto_2022}.  A possible route to addressing these challenges is provided by the paradigm of hybrid quantum-classical variational techniques. In the context of quantum computing, variational quantum algorithms were introduced as an algorithmic framework that may be feasible even when the effects of noise cannot be suppressed to an arbitrarily low level \cite{Peruzzo_2014, Farhi_2014, Farhi_2014_2, Cerezo_2021}. More recently, variational approaches to quantum metrology have emerged as a potential application \cite{Kaubruegger_2019, Zhuang_2019, Koczor_2020, Pezze_2020, Schulte_2020, Yang_2020, Yang_2021, Kaubruegger_2021, Meyer_2021, Ma_2021, Altherr_2021, Xia_2021, Carrasco_2022, Marciniak_2022, Beckey_2022, Zheng_2022, Ho_2022, Liu22}. However, the most efficient use of available entangling resources is not known in most scenarios. Further, the robustness of these techniques to noise is not well understood. This second difficulty is partially due to the breakdown of standard symmetric subspace simulation techniques in the presence of spatially correlated noise. In this paper, we partially address both of these broad challenges.

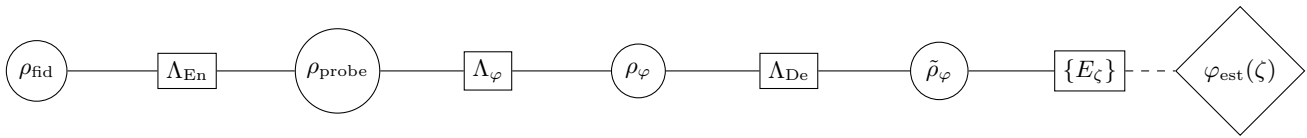
\begin{figure*}
\centering
\begin{tikzpicture}
\node[draw, shape=circle] (fid) at (0,0) {$\rho_{\textrm{fid}}$};
\node[draw, shape=rectangle] (lam_p) at (2,0) {$\Lambda_{\textrm{En}}$};
\node[draw, shape=circle] (prob) at (4,0) {$\rho_{\textrm{probe}}$};
\node[draw, shape=rectangle] (lam_phi) at (6,0) {$\Lambda_{\varphi}$};
\node[draw, shape=circle] (phi) at (8,0) {$\rho_{\varphi}$};
\node[draw, shape=rectangle] (lam_m) at (10,0) {$\Lambda_{\textrm{De}}$};
\node[draw, shape=circle] (meas) at (12,0) {$\tilde{\rho}_{\varphi}$};
\node[draw, shape=rectangle] (povm) at (14,0) {$\{E_{\zeta}\}$};
\node[draw, shape=diamond] (est) at (16,0) {$\varphi_{\textrm{est}}(\zeta)$};

\draw[] (fid) -- (lam_p);
\draw[] (lam_p) -- (prob);
\draw[] (prob) -- (lam_phi);
\draw[] (lam_phi) -- (phi);
\draw[] (phi) -- (lam_m);
\draw[] (lam_m) -- (meas);
\draw[] (meas) -- (povm);
\draw[dashed] (povm) -- (est);
\end{tikzpicture}
\caption{A schematic representation of a general quantum metrology experiment. Time advances from left to right.}
\label{fig:setup}
\end{figure*}

First, in order to optimize the Bayesian mean squared error, we propose a new family of parameterized circuits, called arbitrary-axis twist ansatzes, where the entangling capacity of one-axis twists is effectively optimized with the help of global rotations about arbitrary axes. We find that performance comparable to that of previous approaches can be found with a reduced number of one-axis twists, especially in the few twist regime. Second, we study whether these schemes remain effective in the presence of complex noise. To this end, we utilize a tensor network simulation technique which takes the permutation invariance of the noiseless evolution as a starting point to exactly simulate the protocol in the presence of correlated dephasing noise during the free evolution. We also study the effect of circuit level noise associated with the one-axis twists.

The organization of the paper is as follows. In Sec.~\ref{sec:background}, we review the necessary background. In Sec.~\ref{sec:ansatzes}, we describe our arbitrary-axis twist ansatzes and evaluate their effectiveness.  Finally, in Sec.~\ref{sec:noise}, we study the robustness of our ansatzes to various types of noise. 

\section{Background for quantum metrology}\label{sec:background}
Throughout this paper we consider a system of $N$ qubits or spin-$\frac{1}{2}$ particles. We denote by $X_{j}$, $Y_{j}$, and $Z_{j}$ respectively the Pauli-X, Y, and Z operators on qubit $j$. We will refer to the normalized sum of all Pauli operators in a particular direction, e.g. $J_{z}=\frac{1}{2}\sum_{j}Z_{j}$, as a component of the total angular momentum operator. We denote by $|0\rangle$ or $|1\rangle$, respectively, the $+1$ or $-1$ eigenstate of single qubit Pauli-Z and by $|+\rangle$ or $|-\rangle$, respectively, the $+1$  or $-1$ eigenstate of single qubit Pauli-X. We will also refer to any state of the form $e^{-i\theta_{z}J_{z}}e^{-i\theta_{y}J_{y}}|0\rangle^{\otimes N}$, for any azimuthal angle $\theta_{z}$ and any polar angle $\theta_{y}$, as a spin coherent state.

\subsection{Setting}
The goal of single parameter quantum metrology \cite{Caves_1981, Braunstein_1994, Bollinger_1996, Giovannetti_2004, Giovannetti_2006, Giovannetti_2011, Toth_2014, Degen_2017, Pezze_2018} is to optimally estimate a parameter $\varphi$ associated with some quantum channel $\Lambda_{\varphi}$. Since we are interested in optimal schemes for estimating the parameter, we must specify an objective function by which different schemes may be compared and a feasible set that describes schemes we are willing to consider. The exact form of the objective function is typically dictated by prior knowledge about the value the parameter may take. One method to properly account for prior knowledge (or lack thereof) about the value of $\varphi$ is to treat $\varphi$ as a random variable and pursue a Bayesian statistics approach to the problem. Now knowledge about $\varphi$ before the experiment is preformed can be captured by the prior distribution for this random variable $p(\varphi)$. A useful objective function is then given by the Bayesian mean squared error (BMSE)
\begin{equation}\label{eq:bmse}
\Delta\varphi^{2}\equiv\int_{-\infty}^{\infty}d\varphi\sum_{\zeta}[\varphi-\varphi_{\textrm{est}}(\zeta)]^{2}p(\zeta|\varphi)p(\varphi),
\end{equation}
where the $\{\zeta\}$ are measurement outcomes, $\varphi_{\textrm{est}}(\zeta)$ is an estimator for the value of $\varphi$, and $p(\zeta|\varphi)$ is the probability of measurement outcome $\zeta$ conditioned on the value of $\varphi$. This quantity is also referred to as the averaged estimation error. Such an approach is often called a global approach since it evaluates performance based on estimation error for all values of $\varphi$. In contexts where $\Lambda_{\varphi}[\cdot]=\Lambda_{\varphi+2\pi}[\cdot]$, as we will consider below, this integral is sometimes taken to run for $0$ to $2\pi$ instead. Whether or not this choice makes sense depends on the physical origin of the channel $\Lambda_{\varphi}[\cdot]$. For example, a spin-$\frac{1}{2}$ particle exposed to a strong magnetic field for a fixed amount of time may acquire a relative phase between its spin-up and spin-down states that is equal to $2\pi n+\theta$ for some large integer $n$. This gives the same state as would a relative phase $\theta$ which may arise from a much weaker magnetic field but the underlying physical situation is clearly different. Further, when $\delta\varphi$ is not too large the probability that $|\varphi|>2\pi$ is small and the two approaches should give similar results. In this paper, the integrals in all BMSEs run from $-\infty$ to $\infty$.   

This is in contrast to approaches based on the quantum Fisher information \cite{Braunstein_1994, Giovannetti_2006} which tells us about the possible variance of the estimator at some particular parameter value $\varphi_{0}$. To gain some understanding of the relationship between this averaged estimation error and the Fisher information, note that the BMSE can be written in terms of the variance and bias as
\begin{equation}
\hspace*{-0.5cm}\Delta\varphi^{2}=\int_{-\infty}^{\infty}d\varphi\{\textrm{Var}_{\varphi}[\varphi_{\textrm{est}}]+\textrm{Bias}_{\varphi}[\varphi_{\textrm{est}}]^{2}\}p(\varphi),
\end{equation}
where the variance and bias are defined by
\begin{align}
\textrm{Var}_{\varphi}[\varphi_{\textrm{est}}]&\equiv\sum_{\zeta}(\varphi_{\textrm{est}}(\zeta)-\langle\varphi_{\textrm{est}}\rangle)^{2}p(\zeta|\varphi), \\
\textrm{Bias}_{\varphi}[\varphi_{\textrm{est}}]&\equiv\sum_{\zeta}(\varphi_{\textrm{est}}(\zeta)-\varphi)p(\zeta|\varphi)
\end{align}
and $\langle\cdot\rangle$ denotes the average over measurement outcomes. So this roughly indicates that the Fisher information approach should be appropriate when an estimator can be found that is unbiased or has small bias for all possible values of the parameter $\varphi$. Throughout this paper, we will take the BMSE as the objective function and we will take the prior to be a Gaussian distribution with mean zero and standard deviation $\delta\varphi$.

The basic structure of a quantum metrology experiment is outlined in Fig.~\ref{fig:setup}. The first step is the preparation of some initial fiducial state $\rho_{\textrm{fid}}$. Throughout this paper, we take this state to be the spin coherent state $\rho_{\textrm{fid}}=|+\rangle\langle +|^{\otimes N}$. An ``encoding'' channel, at least partially controlled by the experimenter, is then applied to this state to produce a probe state $\rho_{\textrm{probe}}=\Lambda_{\textrm{En}}[\rho_{\textrm{fid}}]$. By ``partially controlled'' we mean, for example, that the intended probe encoding may consist of unitary dynamics but the quantum channel that is actually realized may be a more general completely positive trace-preserving (CPTP) map due to experimental limitations. This situation is considered in Sec.~\ref{sec:noise}. The probe state then undergoes a ``free evolution" according to the channel $\Lambda_{\varphi}[\cdot]$ that is not controlled by the experimenter and depends on the parameter to be estimated. The resulting state now generally depends on the parameter $\rho_{\varphi}=\Lambda_{\varphi}[\rho_{\textrm{probe}}]$. We will consider channels that reduce to $\Lambda_{\varphi}[\cdot]=e^{-i\varphi J_{z}}\cdot e^{i\varphi J_{z}}$ in the limit that noise is negligible. After the encoding of the parameter $\varphi$, the experimenter can apply a ``decoding" channel $\Lambda_{\textrm{De}}[\cdot]$ to produce the state $\tilde{\rho}_{\varphi}=\Lambda_{\textrm{De}}[\rho_{\varphi}]$. At this point, the state is measured according to some positive operator-valued measure (POVM) with measurement operators $\{E_{\zeta}\}$ giving outcome $\zeta$ with probability $p(\zeta|\varphi)=\textrm{Tr}(\tilde{\rho}_{\varphi}E^{\dag}_{\zeta}E_{\zeta})$. We always take the measurement operators associated with this measurement to be projectors onto the space of states with Hamming weight $w$, i.e. $P_{w}=\sum_{x\ \textrm{s.t.}\ |x|=w}|x\rangle\langle x|$ where $x\in\{0,1\}^{N}$ and $|x|$ is the Hamming weight of $x$, i.e. the number of entries in the bit string $x$ that are equal to one. This constitutes a measurement of the observable $J_{z}$. Note that these operators respect a permutation symmetry and can be implemented experimentally without the ability to detect each particle independently.

The final step in this type of experiment is to process the measurement outcome into an estimate $\varphi_{\textrm{est}}(w)$ of the phase $\varphi$. We take this estimator to be $\varphi_{\textrm{est}}(w)=\frac{a}{2}(N-2w)$ where $a$ is a proportionality constant. The average of this estimator over measurement outcomes is proportional to the expectation value of $J_{z}$ in the state $\tilde{\rho}_{\varphi}$,
\begin{equation}\label{eq:est}
\begin{split}
\langle\varphi_{\textrm{est}}(w)\rangle&\equiv\frac{a}{2}\sum_{w=0}^{N}p(w|\varphi)(N-2w) \\
&=\frac{a}{2}\sum_{w=0}^{N}\textrm{Tr}(P_{w}\tilde{\rho}_{\varphi})((N-w)+(-1)w) \\
&=a\langle J_{z}\rangle,
\end{split}
\end{equation}
where we have used the fact that
\begin{equation}
J_{z}=\frac{1}{2}\sum_{w}(N-w+(-1)w)P_{w}
\end{equation}
since each qubit in $|0\rangle$ increases the Z-angular momentum by $\frac{1}{2}$ and qubit in $|1\rangle$ decreases it by $\frac{1}{2}$. A permutation-invariant measurement operator such as this is known to be optimal in the absence of noise for this type of free evolution \cite{Buzek_1999, Macieszczak_2014}.

The optimal value of the proportionality constant $a$ can be found by first expanding the quadratic form in Eq.~(\ref{eq:bmse}) and performing the averages over the measurement outcomes and the prior distribution to obtain
\begin{equation}
\Delta\varphi^{2}=\delta\varphi^{2}-2a(\varphi\langle J_{z}\rangle)_{\textrm{avg}}+a^{2}\langle J_{z}^{2}\rangle_{\textrm{avg}},
\end{equation}
where $\cdot_{\textrm{avg}}$ denotes the average with respect to the prior $p(\varphi)$. Optimizing this expression with respect to $a$ gives the optimal value of the proportionality constant
\begin{equation}
a_{\textrm{opt}}=\frac{(\varphi\langle J_{z}\rangle)_{\textrm{avg}}}{\langle J_{z}^{2}\rangle_{\textrm{avg}}}.
\end{equation}
This value is fixed by the encoding and decoding procedures and the free evolution. Notably, the value of $a_{\textrm{opt}}$ depends on any noise that may be present. While the resulting estimator is not optimal over all estimators, it is simple to compute and the use of a linear estimator eases comparison to prior work \cite{Kaubruegger_2021, Marciniak_2022} where consideration was also restricted to linear estimators.

The optimal value of the BMSE over all possible probe states, measurements, and estimators can be found via an iterative optimization procedure \cite{Macieszczak_2014, Macieszczak_2013, Chabuda_2020}. In this optimization procedure, first the probe state is fixed and the measurement is optimized for that state. Then the measurement is fixed and the state is optimized for that measurement. This is then repeated until the procedure converges. Since in the absence of noise the optimal strategy is permutation-invariant \cite{Buzek_1999}, this optimization can be restricted to the $N+1$ dimensional permutation-invariant subspace. Unfortunately, however, the output of this optimization does not provide a procedure for realizing this optimal performance with any particular set of resources.

\begin{figure}
(a)
\begin{quantikz}[column sep=5pt, row sep={20pt,between origins}]
\qw & 
\gate[wires=3]{U^{\textrm{En}}_{j}} & \qw \\
\qw & & \qw\\
\qw & & \qw
\end{quantikz}
=
\begin{quantikz}[column sep=5pt, row sep={20pt,between origins}]
\qw & 
\gate[wires=3]{T_{z}(\theta^{(1)}_{j})} & \gate[wires=3]{T_{x}(\theta^{(2)}_{j})} &\gate[wires=3]{R_{x}(\theta^{(3)}_{j})} & \qw \\
\qw & & & & \qw\\
\qw & & & & \qw
\end{quantikz}
\\
(b)
\begin{quantikz}[column sep=5pt, row sep={20pt,between origins}]
\qw & 
\gate[wires=3]{U^{\textrm{De}}_{j}} & \qw \\
\qw & & \qw\\
\qw & & \qw
\end{quantikz}
=
\begin{quantikz}[column sep=5pt, row sep={20pt,between origins}]
\qw & 
 \gate[wires=3]{R_{x}(\phi^{(3)}_{j})}& \gate[wires=3]{T_{x}(\phi^{(2)}_{j})} & \gate[wires=3]{T_{z}(\phi^{(1)}_{j})}& \qw \\
\qw & & & & \qw\\
\qw & & & & \qw
\end{quantikz}
\caption{Illustration of the general form of (a) encoding and (b) decoding layers used in the parity symmetric ansatzes \cite{Kaubruegger_2019, Kaubruegger_2021}. In this diagram, operations further to the left occur first.}
\label{fig:parity}
\end{figure}
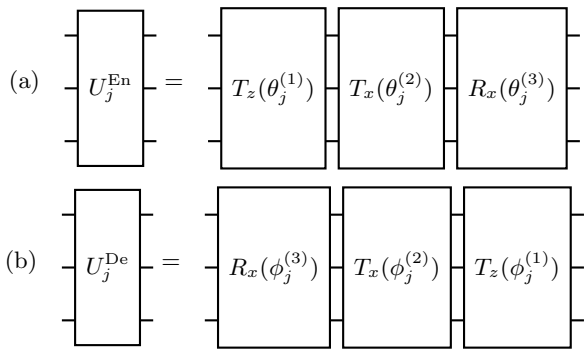

\subsection{Practical resources for quantum metrology}
We now turn the the question of what should constitute the feasible set of strategies. In many experimental setups the available resources include global rotations
\begin{equation}\label{eq:rot}
R_{\mathbf{n}}(\theta)=e^{-i\theta (\mathbf{n}\cdot\mathbf{J})}
\end{equation}
and one-axis twists
\begin{equation}\label{eq:twist}
T_{\mathbf{n}}(\theta)=e^{-i\theta (\mathbf{n}\cdot\mathbf{J})^{2}},
\end{equation}
where $\mathbf{n}$ is a 3 dimensional unit vector and $\mathbf{J}=J_{x}e_{x}+J_{y}e_{y}+J_{z}e_{z}$~\cite{Kitagawa_1993, Henkel_2010, Pupillo_2010, Gil_2014}. Here $e_{\mu}$ is the unit vector directed along the $\mu$-axis. We will frequently use the short hand notations $R_{\mu}(\theta)\equiv R_{e_{\mu}}(\theta)$ and $T_{\mu}(\theta)\equiv T_{e_{\mu}}(\theta)$ for global rotations and one-axis twists along the \emph{x}, \emph{y}, or \emph{z}-axes respectively. Measurements of the total angular momentum of the ensemble along some axis are also commonly possible. Accordingly, we will take the feasible set of strategies to be defined by the set of probe states that can be prepared from $|+\rangle^{\otimes N}$ by a specified sequence of global rotations and one-axis twists and the set of POVMs that can be implemented via a (possibly different) specified sequence of global rotations and one-axis twists ending with a global $\frac{\pi}{2}$ \emph{x}-rotation $R_x(\frac{\pi}{2})$ and followed by a measurement of the \emph{z}-component of the total angular momentum. Note that all of the entanglement used in these schemes comes from the one-axis twists. These are the resources used in twist-untwist protocols \cite{Leibfried_2004, Leibfried_2005, Davis_2016, Frowis_2016, Macri_2016, Nolan_2017, Colombo_2022, Li_2022}. In these protocols, the encoding operation consists of a one-axis twist about the \emph{z}-axis by $\frac{1}{\sqrt{N}}$ followed by a rotation about the \emph{x}-axis by $\frac{\pi}{2}$. The decoding is then the inverse of the encoding operation followed by a final rotation about the \emph{x}-axis by $\frac{\pi}{2}$.

\subsection{Variational metrology and parity symmetric ansatzes}

The basic idea of quantum variational algorithms is to prepare, on the quantum device, a classically parameterized state by applying a series of unitaries that depend on classical parameters to some fiducial quantum state following which some efficient measurement is preformed on the resulting state. The outcome of this measurement, or a series of such measurements, is then used to update the values of the classical parameters. The sequence of unitaries used here is referred to as a parameterized circuit or an ansatz. We will replace both the encoding channel and the decoding channel described above with appropriate parameterized circuits and then numerically find the optimal values for the classical parameters.

We are interested primarily in studying the performance of two sets of parameterized circuits. The first set was introduced originally by a pioneering work of Kaubruegger et al. \cite{Kaubruegger_2019, Kaubruegger_2021}. This family of ansatzes is described in terms of a number of encoding layers $L_{\textrm{En}}$ and a number of decoding layers $L_{\textrm{De}}$ so that the total encoding and decoding circuits can be written $U^{({\textrm{En}/\textrm{De}})}=\prod_{j=1}^{L_{\textrm{En}/\textrm{De}}}U^{(\textrm{En}/\textrm{De})}_{j}$. Each layer is composed of two one-axis twists as defined in Eq.~(\ref{eq:twist}) and a global rotation as defined in Eq.~(\ref{eq:rot}). The encoding and decoding layers respectively have the form
\begin{align}
U^{\textrm{En}}_{j}&=R_{x}(\theta^{(3)}_{j})T_{x}(\theta^{(2)}_{j})T_{z}(\theta^{(1)}_{j}), \\
U^{\textrm{De}}_{j}&=T_{z}(\phi^{(1)}_{j})T_{x}(\phi^{(2)}_{j})R_{x}(\phi^{(3)}_{j}),
\end{align}
as illustrated in Fig.~\ref{fig:parity}. After the decoding unitary is applied, an $R_{x}\left(\frac{\pi}{2}\right)$ rotation is performed followed by a measurement of the \emph{z}-component of total angular momentum as described in Eq.~(\ref{eq:est}). We call these the parity symmetric ansatzes since all gates used in the circuit independently commute with the X-parity operator $P_{X}=\prod_{j}X_{j}$. This property implies that the resulting estimator will be anti-symmetric in $\varphi$ and that the probe state's average angular momentum will be directed along the \emph{x}-axis. The second property is important because spin squeezed states oriented along the \emph{x}-axis are likely to preform well. The member of this family of ansatzes with $L_{\textrm{En}}$ encoding layers and $L_{\textrm{De}}$ decoding layers will be referred to as the $\textrm{PAR}^{2L_{\textrm{En}}}_{2L_{\textrm{De}}}$ ansatz since $2L_{\textrm{En}}$ one-axis twists are used in the encoding and $2L_{\textrm{De}}$ twists are used in the decoding. The parity symmetric ansatzes were found to achieve nearly the optimal estimation performance allowed by quantum mechanics \cite{Macieszczak_2014} when the number of layers is sufficiently large, see for reference Fig.~2 in Ref.~\cite{Kaubruegger_2021}.

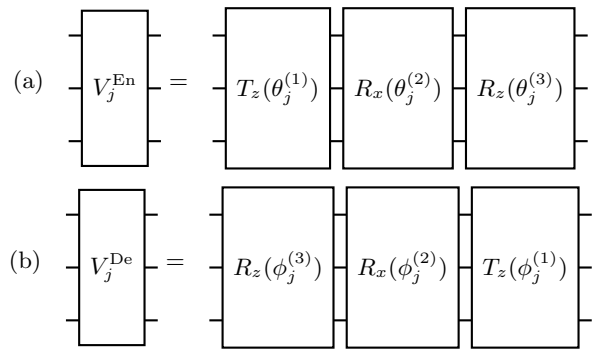
\begin{figure}
(a)
\begin{quantikz}[column sep=5pt, row sep={20pt,between origins}]
\qw & 
\gate[wires=3]{V^{\textrm{En}}_{j}} & \qw \\
\qw & & \qw\\
\qw & & \qw
\end{quantikz}
=
\begin{quantikz}[column sep=5pt, row sep={20pt,between origins}]
\qw & 
\gate[wires=3]{T_{z}(\theta^{(1)}_{j})} & \gate[wires=3]{R_{x}(\theta^{(2)}_{j})} & \gate[wires=3]{R_{z}(\theta^{(3)}_{j})} & \qw \\
\qw & & & & \qw\\
\qw & & & & \qw
\end{quantikz}
\\
(b)
\begin{quantikz}[column sep=5pt, row sep={20pt,between origins}]
\qw & 
\gate[wires=3]{V^{\textrm{De}}_{j}} & \qw \\
\qw & & \qw\\
\qw & & \qw
\end{quantikz}
=
\begin{quantikz}[column sep=5pt, row sep={20pt,between origins}]
\qw & 
 \gate[wires=3]{R_{z}(\phi^{(3)}_{j})} & \gate[wires=3]{R_{x}(\phi^{(2)}_{j})} & \gate[wires=3]{T_{z}(\phi^{(1)}_{j})} & \qw \\
\qw & & & & \qw\\
\qw & & & & \qw
\end{quantikz}
\caption{Illustration of the general form of the unitaries that are appended to the (a) encoding and (b) decoding circuits to increase the number of one-axis twists once at least one is in use in the arbitrary-axis twist ansatzes. In this diagram, time proceeds from left to right.}
\label{fig:twist}
\end{figure}

\begin{figure*}
\begin{overpic}[width=0.46\textwidth]{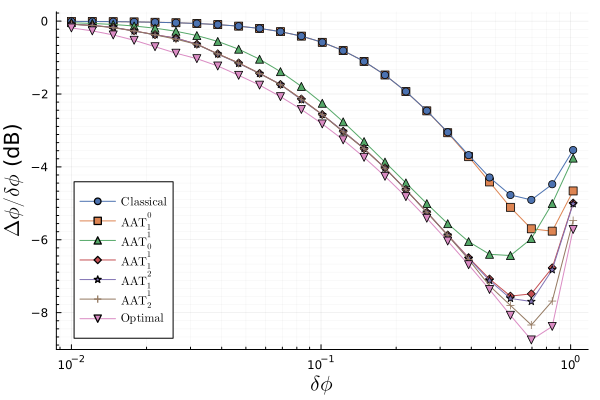}
\put (-1,60) {(a)}
\end{overpic}
\begin{overpic}[width=0.46\textwidth]{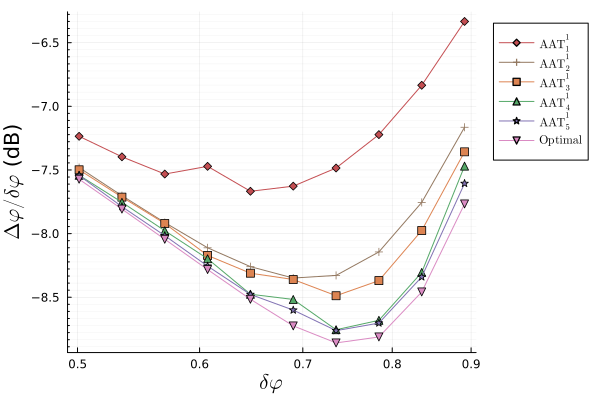}
\put (-1,60) {(b)}
\end{overpic}\\
\begin{overpic}[width=0.46\textwidth]{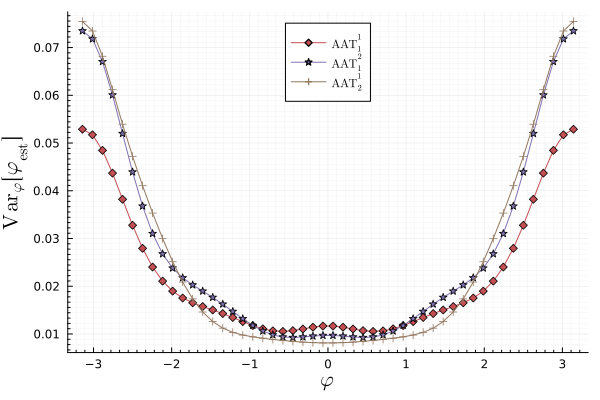}
\put (-1,60) {(c)}
\end{overpic}
\begin{overpic}[width=0.46\textwidth]{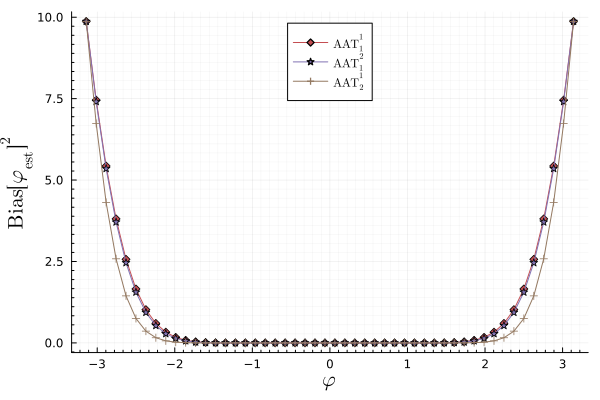}
\put (-1,60) {(d)}
\end{overpic}
\caption{(a) The optimized root Bayesian mean squared error $\Delta\varphi/\delta\varphi$ of various low depth arbitrary-axis twist ansatzes plotted vs. the prior standard deviation $\delta\varphi$. (b) Plot of $\Delta\varphi/\delta\varphi$ vs. $\delta\varphi$ for up to five decoding twists for $0.5\lesssim\delta\varphi\lesssim0.9$. In (c) and (d) respectively, the variances and squared biases of the estimators associated with $\textrm{AAT}^{1}_{1}$, $\textrm{AAT}^{1}_{2}$, and $\textrm{AAT}^{2}_{1}$ are plotted vs. $\varphi$ at $\delta\varphi\approx 0.74$. In all plots $N=30$.}
\label{fig:lowdepth}
\end{figure*}

\section{Arbitrary-Axis Twist Ansatzes}\label{sec:ansatzes}
\subsection{Definition and motivation}

We propose a new family of parameterized circuits for this problem which we refer to as the arbitrary-axis twist ansatzes. The encoding and decoding unitaries are alternations of arbitrary global rotations and one-axis twists about the \emph{z}-axis. The unitaries always begin and end with an arbitrary rotation so that they effectively implement a series of one-axis twists and global rotations about arbitrary axes followed by an additional rotation. We will label these ansatzes by the number of one-axis twists used in the encoding and the number of one-axis twists used in the decoding. The arbitrary-axis twist ansatz using $n_{\textrm{En}}$ one-axis twists in the encoding and $n_{\textrm{De}}$ one-axis twists in the decoding will be referred to as $\textrm{AAT}^{n_{\textrm{En}}}_{n_{\textrm{De}}}$. For example, $\textrm{AAT}^{1}_{1}$ can be thought of as a family of generalized twist-untwist protocols. The encoding and decoding unitaries used in $\textrm{AAT}^{1}_{1}$ are
\begin{align}
\hspace*{-0.2cm}V^{(\textrm{En})}_{1}&=R_{z}(\theta^{(5)}_{1})R_{x}(\theta^{(4)}_{1})T_{z}(\theta^{(3)}_{1})R_{z}(\theta^{(2)}_{1})R_{y}(\theta^{(1)}_{1}), \\
\hspace*{-0.2cm}V^{(\textrm{De})}_{1}&=R_{x}(\phi^{(1)}_{1})R_{z}(\phi^{(2)}_{1})T_{z}(\phi^{(3)}_{1})R_{x}(\phi^{(4)}_{1})R_{z}(\phi^{(5)}_{1}).
\end{align}
As for the parity symmetric ansatzes, we also include a final $R_{x}(\frac{\pi}{2})$ rotation at the end of the decoding immediately prior to the $J_{z}$ measurement. Each additional one-axis twists requires three new parameters. In particular, adding an additional encoding or decoding one-axis twist means appending the following unitaries to the end (beginning) of the encoding (decoding) circuit,
\begin{align}
V^{(\textrm{En})}_{j}&=R_{z}(\theta^{(3)}_{j})R_{x}(\theta^{(2)}_{j})T_{z}(\theta^{(1)}_{j})\ , \\
V^{(\textrm{De})}_{j}&=T_{z}(\phi^{(1)}_{j})R_{x}(\phi^{(2)}_{j})R_{z}(\phi^{(3)}_{j})\ .
\end{align}
The form of these appended unitaries is illustrated in Fig.~\ref{fig:twist}. Altogether $\textrm{AAT}^{n_{\textrm{En}}}_{n_{\textrm{De}}}$ contains $4+3(n_{\textrm{En}}+n_{\textrm{De}})$ circuit parameters.

We note that using sequential rotations about only 2 axes, rather than the standard 3 rotations by Euler angles, is sufficient to make the global rotations arbitrary. To understand why, notice that we can write a one-axis twist flanked by two different arbitrary global rotations as
\begin{equation}
\begin{split}
&R_{z}(\gamma')R_{x}(\beta')R_{z}(\alpha')T_{z}(\omega)R_{z}(\gamma)R_{x}(\beta)R_{z}(\alpha) \\
&=R_{z}(\gamma')R_{x}(\beta')T_{z}(\omega)R_{z}(\alpha'+\gamma)R_{x}(\beta)R_{z}(\alpha).
\end{split}
\end{equation}
So the number of rotations can be reduced by one. This can be repeated for each set of three rotations. Since the fiducial state is taken to be $|+\rangle$ a parameter can be removed from the first rotation by using the Euler decomposition $R_{z}(\gamma)R_{y}(\beta)R_{x}(\alpha)$ and dropping the \emph{x}-rotation. Similarly, since the final measurement is effectively a measurement of $J_{y}$, a parameter can be removed from the last global rotation by using the Euler decomposition $R_{y}(\gamma)R_{x}(\beta)R_{z}(\alpha)$ and dropping the last rotation.

Notably, while it utilizes only the one-axis twists about the fixed \emph{z}-axis, $\textrm{AAT}^{n_{\textrm{En}}}_{n_{\textrm{De}}}$ is the most general ansatz composed of one-axis twists and global rotations with $n_{\textrm{En}}$ twists used in the encoding and $n_\textrm{De}$ used in the decoding. This is because the rotations can effectively change the axis of the one-axis twists, so that this family of ansatzes is universal to generate the set of unitaries which are composed of a sequence of one-axis twists about arbitrary axes interspersed with any global rotations. Due to the generality of the arbitrary-axis twist ansatzes, $\textrm{AAT}^{n_{\textrm{En}}}_{n_{\textrm{De}}}$ will preform at least as well as $\textrm{PAR}^{n_{\textrm{En}}}_{n_{\textrm{De}}}$. However, $\textrm{PAR}^{2n_{\textrm{En}}}_{2n_{\textrm{De}}}$ has $3(n_{\textrm{En}}+n_{\textrm{De}})$ parameters while $\textrm{AAT}^{2n_{\textrm{En}}}_{2n_{\textrm{De}}}$ has $4+6(n_{\textrm{En}}+n_{\textrm{De}})$ parameters. Thus, comparisons between the members of these two families of ansatzes with a similar number of parameters represent a comparison between the power of increased freedom in rotations and increased entanglement. 

\subsection{Noiseless performance}

\begin{figure*}
\begin{overpic}[width=0.49\textwidth]{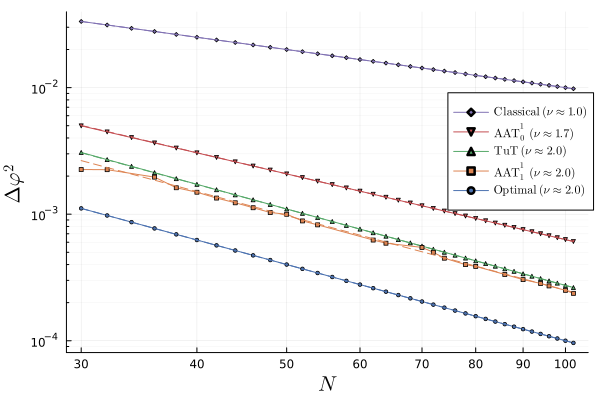}
\put (-1,60) {(a)}
\end{overpic}
\begin{overpic}[width=0.49\textwidth]{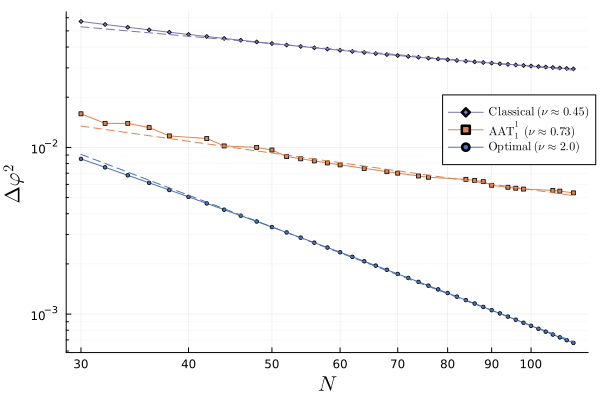}
\put (-1,60) {(b)}
\end{overpic}
\caption{Plots of the shifted Bayesian mean squared error $\Delta\tilde{\varphi}^{2}$ vs. system size $N$ for prior standard deviations of (a) $\delta\varphi=10^{-3}$ and (b) $\delta\varphi=0.7$. The markers on the solid curves correspond to the performance of the strategies the dashed lines are the fitted curves. In (a) the optimal, $\textrm{AAT}^{1}_{2}$, twist-untwist (TuT) strategies are all fit the curves with exponents $\nu\approx 2.0$, the $\textrm{AAT}^{1}_{0}$ is fit to a curve with $\nu \approx1.7$, and the classical strategy is fit to a curve with $\nu\approx1.0$. In (b) the optimal strategy is again fit to a curve with $\nu\approx2.0$, the $\textrm{AAT}^{1}_{1}$ strategy is fit to a curve with $\nu\approx0.73$, and the classical strategy is fit to a curve with $\nu\approx0.45$.}
\label{fig:fit}
\end{figure*}

To analyze the performance of these ansatzes we begin by numerically optimizing the circuit parameters for several of the low depth arbitrary-axis twist ansatzes over a wide range of values of the prior variance $\delta\varphi$. While the simulation of the quantum evolution is efficient due to the permutation invariance in the ideal setting (and low-bond dimension matrix product state representations in the noisy setting), the classical optimization of the circuit parameters can be a major limitation as the number of circuit parameters is increased. Accordingly, we consider a system size of $N=30$ throughout this section. See App.~\ref{sec:numerics} for details on how the numerical optimization was preformed. In Fig.~\ref{fig:lowdepth}a, we plot the Bayesian mean squared error we obtain vs. the prior standard deviation. In particular, we compare $\textrm{AAT}^{0}_{1}$, $\textrm{AAT}^{1}_{0}$, $\textrm{AAT}^{1}_{1}$, $\textrm{AAT}^{1}_{2}$, and $\textrm{AAT}^{2}_{1}$. For reference, we also include curves corresponding to the best entanglement free (i.e. classical) approach and the optimal approach allowed by quantum mechanics~\cite{Macieszczak_2014}. We observe a first dramatic improvement in performance when moving from the classical approach to $\textrm{AAT}^{1}_{0}$. A second large improvement occurs when moving from $\textrm{AAT}^{1}_{0}$ to $\textrm{AAT}^{1}_{1}$. Finally we notice another significant improvement when moving from $\textrm{AAT}^{1}_{1}$ to $\textrm{AAT}^{1}_{2}$. This is indicates that, once there is a single encoding twist, the most promising place to look for improvement is in adding twists to the decoding. This also makes sense in light of previous observations that the optimal probe states are well approximated by spin squeezed states in this case~\cite{Macieszczak_2014}.

We now focus on the regime where the largest improvement over the classical approach is possible, i.e. $0.5\lesssim\delta\varphi\lesssim 0.9$. In this region, we examine how increasing the depth of the decoding circuit effects the BMSE. A comparison for $\textrm{AAT}^{1}_{1}$, $\textrm{AAT}^{1}_{3}$, $\textrm{AAT}^{1}_{4}$, and $\textrm{AAT}^{1}_{5}$ is shown in Fig.~\ref{fig:lowdepth}b. We observe that as the circuit depth is increased the optimized estimators approach the optimal value of the BMSE. However, we also note that increasing the circuit depth appears to provide diminishing returns as the optimal is approached. 
While this is to some extent inevitable due to the existence of a lower bound, we observe strikingly little improvement in the transition from $\textrm{AAT}^{1}_{4}$ to $\textrm{AAT}^{1}_{5}$. Due to these diminishing returns it is reasonable to consider circuits with only a few one-axis twists in at least some cases.

We next examine the origin of the improvement of $\textrm{AAT}^{1}_{2}$ over $\textrm{AAT}^{1}_{1}$. To do this we plot, in Fig.~\ref{fig:lowdepth}c, the variance of the optimized estimators at $\delta\varphi\approx0.74$ over a range of values of $\varphi$ for $\textrm{AAT}^{1}_{1}$, $\textrm{AAT}^{2}_{1}$, and $\textrm{AAT}^{1}_{2}$. We note that neither $\textrm{AAT}^{1}_{2}$ nor $\textrm{AAT}^{2}_{1}$ exhibit dramatic improvement over $\textrm{AAT}^{1}_{1}$. In Fig.~\ref{fig:lowdepth}d, we plot the bias squared of the optimized estimators over measurement outcomes at the same value of $\delta\varphi$ and same range of $\varphi$ for $\textrm{AAT}^{1}_{1}$, $\textrm{AAT}^{2}_{1}$, and $\textrm{AAT}^{1}_{2}$. We observe that the estimator associated with $\textrm{AAT}^{1}_{2}$ displays a larger region of low bias than the other two ansatzes. This is in line with previous observations that increasing the depth of the decoding circuit primarily serves to decrease bias~\cite{Marciniak_2022}. It is also not surprising that a decrease in bias should be the most effective way to obtain a smaller BMSE since we are examining the case where the prior variance is fairly large.

\subsection{System size dependence and scaling}

Next, we examine how the performance of our ansatz compares to that of the optimal and classical strategies as the system size is increased with an emphasis on scaling. The quantum van Trees~\cite{vanTrees_1968, Gill_1995, Paris_2009, Tsang_2020, Kaubruegger_2021}, or Bayesian quantum Cram\'{e}r-Rao, inequality states that
\begin{equation}
\Delta\varphi^{2}\geq\frac{1}{F_{Q}+F_{P}},
\end{equation}
where $F_{Q}$ is the quantum Fisher information averaged over the prior distribution and $F_{P}$ is the Fisher information of the prior distribution. In the absence of noise, the quantum Fisher information is independent of the value of $\varphi$ and thus this averaged Fisher information satisfies
\begin{equation}
F_{Q}\leq4(\textrm{Tr}(J_{z}^{2}\rho_{\textrm{probe}})-\textrm{Tr}(J_{z}\rho_{\textrm{probe}})^{2})\leq N^{2}.
\end{equation}
The Fisher information of a Gaussian distribution is given by
\begin{equation}
F_{P}=\frac{1}{\delta\varphi^{2}}.
\end{equation}
Motivated by this, we simplify the curve fitting process by considering a shifted version of the BMSE defined by
\begin{equation}
(\Delta\varphi')^{-2}\equiv\Delta\varphi^{-2}-\delta\varphi^{-2}.
\end{equation}
The fitting is further simplified by subtracting out the excess mean-squared error associated with phase slips, i.e. times when the $\phi<\-\pi$ or $\phi>\pi$. When this happens the mean-squared error is approximately $4\pi^2$. Further, the probability of this occurring is $1-\textrm{erf}\left(\frac{\pi}{\sqrt{2}\delta\varphi}\right)$. Accordingly, we further transform to 
\begin{equation}
\Delta\tilde{\varphi}^{2}\equiv\Delta\varphi'^{2}-4\pi^{2}\left(1-\textrm{erf}\left(\frac{\pi}{\sqrt{2}\delta\varphi}\right)\right).
\end{equation}
In order to study the scaling, we then fit $\Delta\tilde{\varphi}$ to a curve of the form
\begin{equation}
f(N)=\frac{\alpha}{N^{\nu}},
\end{equation}
where $\alpha$ and $\nu$ are fitting parameters. When performing the fitting we exclude points where optimization becomes stuck in a local minimum.

The results are shown in Fig.~\ref{fig:fit} for prior standard deviations of (a) $\delta\varphi=10^{-3}$ and (b) $\delta\varphi=0.7$. The markers on the solid line indicate the performance of the strategy at a particular system size and the dashed lines indicate the fittings. In Fig.~\ref{fig:fit} (a), where the prior standard deviation is small, the BMSE is dominated by variance and the optimal and classical curves respectively exhibit the familiar Heisenberg $\nu\approx2$ and standard quantum limit $\nu\approx1$ scaling. Our $\textrm{AAT}^{1}_{0}$ shows a scaling in good agreement with that expected of spin squeezed states $\nu\approx1.7\approx\frac{5}{3}$~\cite{Kitagawa_1993}. Our $\textrm{AAT}^{1}_{1}$ ansatz exhibits Heisenberg scaling for large enough system sizes. For reference, we also plot the performance of the twist-untwist (TuT) protocol described in~\cite{Davis_2016}. The twist-untwist protocol exhibits comparable but slightly worse performance than our $\textrm{AAT}^{1}_{1}$  ansatz which makes sense given that the $\textrm{AAT}^{1}_{1}$ ansatzes are generalized twist-untwist protocols. In Fig.~\ref{fig:fit} (b), where the prior standard deviation is much larger, we find that the exponents of the optimal strategy continues to exhibit Heisenberg scaling as expected from the $\pi$-corrected Heisenberg limit~\cite{Gorecki_2020}. However, while the optimal scaling is independent of the prior standard deviation, the variational strategies only achieve the optimal performance at large prior standard deviation if a large number of layers are used. For this reason, we see a decrease in the scaling exponent when the prior standard deviation is increased but the number of layers is kept small. The scaling of the classical strategy decreases to $\nu\approx0.45$. This decrease may be due to the restriction that $\langle\varphi_{\textrm{est}}\rangle\propto\langle J_{z}\rangle$. Our $\textrm{AAT}^{1}_{1}$ ansatz exhibits an intermediate scaling with $\nu\approx0.73$. From these observations we infer that even our low depth arbitrary-axis twist ansatzes can achieve better scaling than the corresponding classical strategies over a wide range of values of prior standard deviation.

\subsection{Comparison to the previously considered parity symmetric ansatzes}
\begin{figure}
\includegraphics[width=0.49\textwidth]{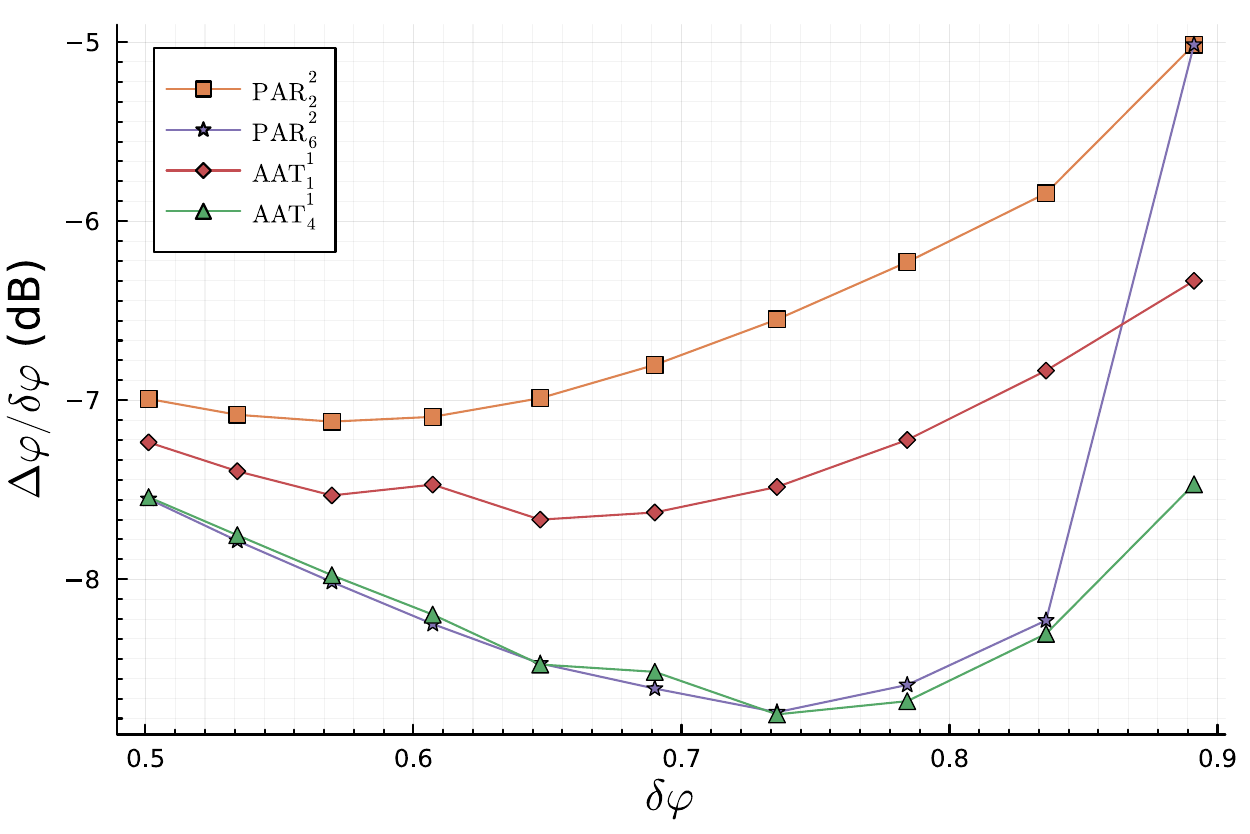}
\caption{Plot of the root Bayesian mean squared error $\Delta\varphi/\delta\varphi$ vs. prior variance $\delta\varphi$ for the optimized ansatzes $\textrm{PAR}^{2}_{2}$, $\textrm{PAR}^{2}_{6}$, $\textrm{AAT}^{1}_{1}$, and $\textrm{AAT}^{1}_{4}$ in the $0.5\lesssim\delta\varphi\lesssim0.9$ regime for $N=30$. $\textrm{AAT}^{1}_{1}$ preforms better than $\textrm{PAR}^{2}_{2}$ and $\textrm{AAT}^{1}_{4}$ preforms about as well as $\textrm{PAR}^{2}_{6}$. We note that the rightmost $\textrm{PAR}^{2}_{6}$ point is trapped in a local minima in this numerical setting, but can be resolved by a different initialization, for instance. See App.~\ref{sec:numerics} and \ref{sec:compare} for details on how the optimization was preformed and values of the Bayesian mean squared error that can be obtained with this ansatz and prior variance via other optimization methods.}
\label{fig:ansatz_comp}
\end{figure}

Finally, we compare the performance of the optimized arbitrary-axis twist estimators to that of the optimized parity symmetric estimators with low and medium depth decoding circuits introduced in~\cite{Kaubruegger_2019, Kaubruegger_2021, Marciniak_2022}. The results of this comparison are shown in Fig.~\ref{fig:ansatz_comp}. We find that $\textrm{AAT}^{1}_{1}$ achieves a smaller BMSE than $\textrm{PAR}^{2}_{2}$ despite using only half as many one-axis twists. We also observe that $\textrm{AAT}^{1}_{4}$ performs comparably to $\textrm{PAR}^{2}_{6}$ despite the former using only 5 one-axis twists and the latter using 8. We also have some evidence that the optimization of the arbitrary-axis twist ansatzes seems to be eased for numerical optimization. In particular, we find that arbitrary-axis twist ansatzes may be less likely to fail due to local minima compared to the parity symmetric ansatzes with a similar number of parameters. See App.~\ref{sec:compare} for more details about the comparison of the performance of these two families of ansatzes and App.~\ref{sec:time} for information about the total twisting times associated with the arbitrary-axis twist strategies.

\section{Tensor network simulation for noisy quantum metrology}\label{sec:noise}

As the effect of single qubit dephasing noise during the free evolution in variational quantum metrology has been studied previously in both the Bayesian~\cite{Kaubruegger_2021} and Fisher information settings~\cite{Koczor_2020, Yang_2020, Meyer_2021, Ma_2021, Zheng_2022, Liu22}, here we focus on the effect of more complex noise such as correlated dephasing models. We also study the effect of circuit level noise during the encoding and decoding circuits.

\subsection{Correlated dephasing}\label{sec:corr}

In realistic scenarios, the signal $\varphi$ to be detected may not be constant over all qubits. We consider the case where the signal $\varphi_{j}$ at each site can be broken up as $\varphi_{j}=\varphi+r_{j}$ where $\varphi$ is a constant piece at each site and $r_{j}$ is a Gaussian random variable with mean zero. In particular, we take the spatial correlation functions associated with this random variable to be $\overline{r_{j}r_{k}}\equiv C_{j,k}$, where $\overline{\cdot}$ indicates average over the Gaussian distribution that the $r_{j}$ are drawn from and we will assume, as a first nontrivial example, that $C_{j,k}=0$ if $|j-k|>1$. The physical content of this is that we allow for correlations but only between nearest neighbor sites in 1D. Such an evolution can be thought of as a global \emph{z}-rotation followed by a correlated dephasing channel. The effective noise channel is then
\begin{align}\label{eq:noise}
\begin{split}
&\mathcal{N}[\rho]=\overline{e^{-i\sum_{j}\frac{r_{j}}{2}Z_{j}}\rho e^{i\sum_{j}\frac{r_{j}}{2}Z_{j}}} \\
&=\sum_{m,n\in\{0,1\}^{N}}\rho_{m,n}\overline{e^{-\frac{i}{2}(\sum_{j,k}(-1)^{m_{j}}r_{j}-(-1)^{n_{k}}r_{k})}}|m\rangle\langle n| \\
&= \sum_{m,n\in\{0,1\}^{N}}\rho_{m,n}e^{-\frac{1}{8}\sum_{j,k}((-1)^{m_{j}}-(-1)^{n_{j}})C_{j,k}((-1)^{m_{k}}-(-1)^{n_{k}})}|m\rangle\langle n|,
\end{split}
\end{align}
where in the last line we explicitly preformed the multivariate Gaussian integration. Such a noise model may be relevant to cavity QED based experiments in which the spins are arranged in a quasi one-dimensional lattice~\cite{Lee_2014, Kawasaki_2019, Braverman_2019, Colombo_2022, Li_2022}.

Up until this point, all dynamics have been restricted to the symmetric subspace, i.e. the $\frac{N}{2}(\frac{N}{2}+1)$ eigenspace of $J_{x}^{2}+J_{y}^{2}+J_{z}^{2}$. This subspace has dimension $N+1$ allowing for efficient simulation. This is a result of the initial states, unitary dynamics, and measurements respecting a permutation symmetry. However, this noise model breaks the permutation symmetry. It has been shown by Chabuda et al. in~\cite{Chabuda_2020} that this noise model nevertheless has an efficient tensor network representation, and a tensor network contraction was utilized to solve recursively for the optimal Fisher information associated with this noisy free evolution. On the other hand, we propose to compile symmetric subspace techniques to tensor network constructions for permutation-invariant states and one-axis twists and combine them with the above construction of correlated dephasing, so that the correlated noise effects during the free evolution of metrology protocols are simulated efficiently.

\subsection{Matrix product states and operators for permutation-invariant states and one-axis twists}\label{sec:tensor}
We now review some basic facts about matrix product states (MPSs) and operators (MPOs) \cite{Perez-Garcia_2007, Orus_2014, Bridgeman_2017, Cirac_2021} and efficient representations of permutation-invariant states and one-axis twists. An MPS representation of a state $|\psi\rangle$ is a representation of the form
\begin{equation}
|\psi\rangle=\sum_{x\in\{0,1\}^{N}}\prod_{j}A^{(x_{j})}[j]|x\rangle,
\end{equation}
where the $\{A^{(0)}[j],A^{(1)}[j]\}$ are matrices, except when $j=1$ or $j=N$ in which case they are respectively row and column vectors. The label $x_{j}$ is called the physical index. The largest dimension of any of the matrices $A^{(s)}[j]$ is referred to as the bond dimension of the representation. Any state of $N$ qubits can be put into an MPS form with a bond dimension of $2^{\lceil\frac{N}{2}\rceil}$.

A natural extension of the MPS construction is the MPO. An MPO representation of an operator $\mathcal{O}$ is a representation of the form
\begin{equation}
\mathcal{O}=\sum_{x^{L},x^{R}\in\{0,1\}^{N}}\prod_{j}A^{(x^{L}_{j},x^{R}_{j})}[j]|x^{L}\rangle\langle x^{R}|.
\end{equation}
The labels $x^{L}_{j}$ and $x^{R}_{j}$ are the called physical indices and the matrix indices are referred to as virtual indices. All operators on $N$ qubits can be put into the form of an MPO with bond dimension $4^{\lceil\frac{N}{2}\rceil}$.

An MPS representation for the state that results from acting an operator with a known MPO representation on a state with a known MPS representation can be determined by contracting the physical indices of the MPO associated with $x^{R}$ with the physical indices of the MPS. If the initial state MPS representation had bond dimension $d_{s}$ with tensors composed of $\{A^{(x_{j})}[j]\}$ and the MPO had bond dimension $d_{o}$ with tensors composed of $\{B^{(x^{L}_{k},x^{R}_{k})}[j]\}$, then the MPS representation for the final state has bond dimension $d_{s}d_{o}$ with
\begin{equation}
A^{(x_{j})}[j]=\sum_{s}B^{(x_{j},s)}[j]\otimes A^{(s)}[j].
\end{equation}
This allows for the simulation of time evolution by contracting an MPO representation of the time evolution operator with an MPS representation of some initial state. If the bond dimension of the state remains polynomial at all times, then expectation values of operators that have polynomial bond dimension MPO representations can be computed with polynomial cost.

An MPS representation with bond dimension $\lceil\frac{N}{2}\rceil+1$ can always be constructed for a permutation-invariant state as follows. Any permutation-invariant state of $N$ qubits can be written as
\begin{equation}
|\psi\rangle=\sum_{x\in\{0,1\}^{N}}c_{|x|}|x\rangle,
\end{equation}
where the coefficients $c_{|x|}$ depend only on the Hamming weight of $x$. The matrices used in the matrix product state representation at a site $j<\lceil\frac{N+1}{2}\rceil$ have dimension $j\times (j+1)$ and elements given by
\begin{align}
(A_{S}^{(0)}[j])_{m,n}&=\delta_{m,n}, \\
(A_{S}^{(1)}[j])_{m,n}&=\delta_{m,n-1}.
\end{align}
The matrices associated with site $N-j+1$ with $j>\lceil\frac{N+1}{2}\rceil$ are just the transpose of those associated with site $j$. The remaining site $j=\lceil\frac{N+1}{2}\rceil$ has matrices
\begin{align}
(A_{S}^{(0)}[\lceil (N+1)/2\rceil])_{m,n}&=c_{m+n}, \\
(A_{S}^{(1)}[\lceil (N+1)/2\rceil])_{m,n}&=c_{m+n+1}.
\end{align}
Analogously, all permutation-invariant operators have a MPO representation with bond dimension $O(N^{3})$ as described in App.~\ref{sec:mpo}.

We now turn to constructing MPOs for one-axis twists. We focus on the one-axis twist about the \emph{z}-axis but similar results hold for any operator that is permutation-invariant and can be diagonalized by single qubit operations. A one-axis twist about the \emph{z}-axis can be written as
\begin{equation}
T_{z}(\theta)=\sum_{x\in\{0,1\}^{N}}e^{-\frac{i}{4}(N-2|x|)^{2}}|x\rangle\langle x|.
\end{equation}
The key observation is that, as was the case for the permutation-invariant state, the coefficients depend only on the Hamming weight. The MPO matrices for a site $j\neq\lceil\frac{N+1}{2}\rceil$ are then given by, $A^{(0,0)}[j]=A_{S}^{(0)}[j]$, $A^{(1,1)}[j]=A_{S}^{(1)}[j]$, and $A^{(0,1)}[j]=A^{(1,0)}[j]=0$. At site $\lceil\frac{N+1}{2}\rceil$ the matrices are given by
\begin{align}
(A^{(0,0)}[\lceil (N+1)/2\rceil])_{m,n}&=e^{-\frac{i}{4}(N-2(m+n))^{2}}, \\
(A^{(1,1)}[\lceil (N+1)/2\rceil])_{m,n}&=e^{-\frac{i}{4}(N-2(m+n+1))^{2}},
\end{align}
and $A^{(0,1)}[\lceil (N+1)/2\rceil]=A^{(1,0)}[\lceil (N+1)/2\rceil]=0$.

The other unitary element of our ansatzes, the global rotations, can be given a bond dimension one representation. For example, the matrices at each site associated with an \emph{x}-rotation are
\begin{equation}
A^{(\mu,\nu)}[j]=\langle\mu|e^{-i\frac{\theta}{2} X}|\nu\rangle, 
\end{equation}
where $\mu,\nu\in\{0,1\}$. Accordingly, when used in time evolution, the global rotations do not cause the bond dimension of the state to increase. This is a result of the fact that the global rotations do not involve interactions between spins. 

Finally, we will need MPO constructions for the operators $J_{z}$ and $J_{z}^{2}$ since we must evaluate expectation values of these operators. The operator $J_{z}$ can be given a bond dimension two representation. At sites away from the boundary $j\neq 1,N$, the matrices are given by
\begin{equation}
A_{J_{z}}^{(\mu,\nu)}[j] = \begin{pmatrix} 1 & (-1)^{\mu}\frac{1}{2} \\
0 & 1 \end{pmatrix}\delta_{\mu,\nu}.
\end{equation}
At $j=1$ and $j=N$, the matrices are respectively $A_{J_{z}}^{(\mu,\nu)}[1]=\begin{pmatrix}1 & (-1)^{\mu}\frac{1}{2}\end{pmatrix}\delta_{\mu,\nu}$ and $A_{J_{z}}^{(\mu,\nu)}[N]=\begin{pmatrix} (-1)^{\mu}\frac{1}{2} \\ 1 \end{pmatrix}\delta_{\mu,\nu}$. The operator $J_{z}^{2}$ can be given a bond dimension three representation. At sites away from the boundary it has matrices given by
\begin{equation}
A_{J_{z}^{2}}^{(\mu,\nu)}[j]=\begin{pmatrix} 1 & (-1)^{\mu}\frac{1}{2} & \frac{1}{4} \\
0 & 1 & (-1)^{\mu} \\
0 & 0 & 1 \end{pmatrix}\delta_{\mu,\nu}.
\end{equation}
At the boundary sites, the matrices are given by $A_{J_{z}^{2}}^{(\mu,\nu)}[1]=\begin{pmatrix}1 & (-1)^{\mu}\frac{1}{2} &
\frac{1}{4} \end{pmatrix}\delta_{\mu,\nu}$ and $A_{J_{z}^{2}}^{(\mu,\nu)}[N]=\begin{pmatrix} \frac{1}{4} \\
(-1)^{\mu}\frac{1}{2} \\ 1\end{pmatrix}\delta_{\mu,\nu}$.

\subsection{Simulation algorithm}

\begin{figure}
\includegraphics[width=0.49\textwidth]{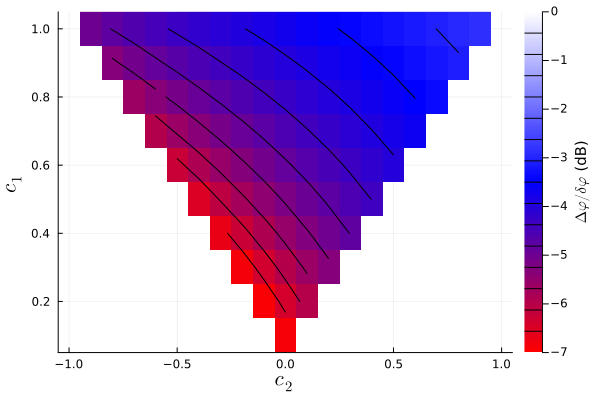}
\caption{Reduction of the root Bayesian mean squared error $\Delta\varphi/\delta\varphi$ for the optimized $\textrm{AAT}^{1}_{1}$ ansatz under a correlated dephasing channel of Eq.~(\ref{eq:noise}) during a free evolution, when $N=30$ and $\delta\varphi\approx 0.74$. The correlated noise is characterized in terms of the variance of $r_{j}$ at each site $c_{1}$ and the covariances at adjacent sites $c_{2}$. The curves of constant $\Delta\varphi/\delta\varphi$ indicate that correlated dephasing is not much worse than uncorrelated and anti-correlated dephasing is more favorable than uncorrelated.}
\label{fig:free}
\end{figure}

We use these constructions as a basis for simulations of metrology experiments in the presence of noise that breaks the permutation symmetry. If 1D geometrically local noise, for example the correlated dephasing described above, occurs during the free evolution, we introduce a simulation strategy that makes use of both symmetric subspace simulation and matrix product constructions. 

Initially, the first three stages in Fig.~\ref{fig:setup} are simulated in the symmetric subspace so that an $(N+1)\times(N+1)$ symmetric subspace representation of $\rho_{\textrm{probe}}$ is obtained. Then this is converted to an MPO representation using the construction for permutation-invariant states described in the last section.

Second, if the only entangling resources used are one-axis twists and the number of one-axis twists is sufficiently small, the decoding, i.e. stages 6 and 7 in Fig.~\ref{fig:setup}, can be dealt with in two ways. The first method is to map each one-axis to an MPO as described in the last section. The MPOs for all one-axis twists and global rotations used in the decoding can then be contracted to obtain an MPO for the time evolution operator associated with the decoding. Alternatively, if the number of one-axis twists is large enough that the first procedure would lead to an intractable bond dimension or entangling resources other than one-axis twists are used, the decoding unitary may be compiled into a single permutation-invariant unitary as described in App.~\ref{sec:compile}. If the depth of the circuit is $\ell$, then the cost of this compilation is $O(\ell N^{7})$ in time. Note that it is not sufficient to work only in the symmetric subspace when compiling the decoding unitary as it will be acting on a state that lies outside of the symmetric subspace. The resulting permutation-invariant unitary can then be mapped to an MPO with bond dimension $O(N^{3})$ via a construction analogous to the one for permutation-invariant states as noted above and in App.~\ref{sec:mpo}. The first approach has the advantage of allowing the compiling step to be skipped, while the advantage of the second approach is that by construction the bond dimension of the resulting MPO is guaranteed to be polynomial in system size. In this paper, we utilize the former approach in which the MPOs for individual one-axis twists are independently constructed.

Next, an MPO is constructed for the superoperator associated with the free evolution, i.e. stages 4 and 5 in Fig.~\ref{fig:setup}. The correlated dephasing model described above in Eq.~\ref{eq:noise} has an MPO representation with bond dimension $2$~\cite{Chabuda_2020}. Since this noise commutes with the unitary part of the free evolution we consider and the unitary part is a global rotation, the entire free evolution superoperator can be given a bond dimension 3 representation.

Finally, the MPOs obtained in these three steps can then be contracted to obtain an MPO of the state immediately before measurement. If the number of one-axis twists used in the decoding does not scale with system size and the MPO for the free evolution superoperator has polynomial bond dimension, either decoding simulation strategy gives an algorithm with cost polynomial in system size. If the number of gates used in the decoding scales polynomially with system size then the compiling strategy continues to give a polynomial cost algorithm.

\subsection{Robustness to correlated noise}

\begin{figure*}
\begin{overpic}[width=0.49\textwidth]{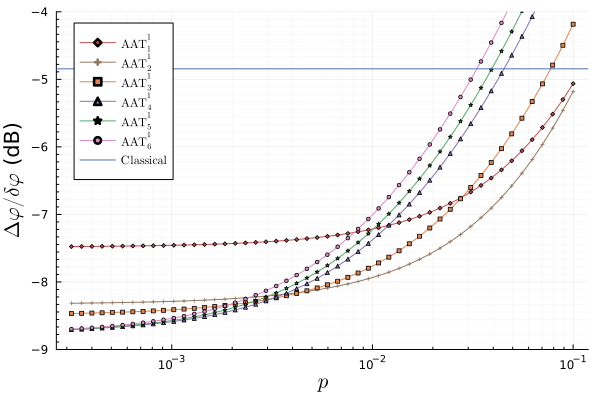}
\put (-1,60) {(a)}
\end{overpic}
\begin{overpic}[width=0.49\textwidth]{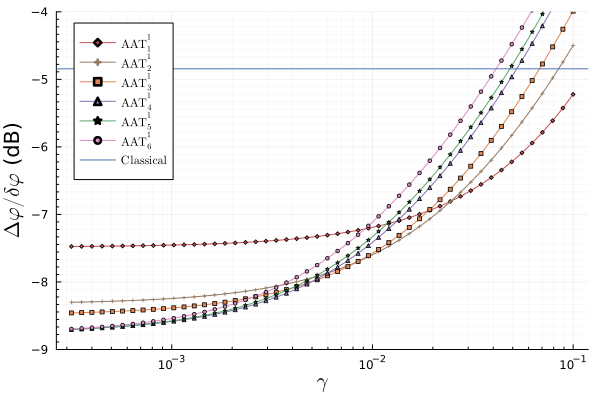}
\put (-1,60) {(b)} 
\end{overpic} \\
\begin{overpic}[width=0.49\textwidth]{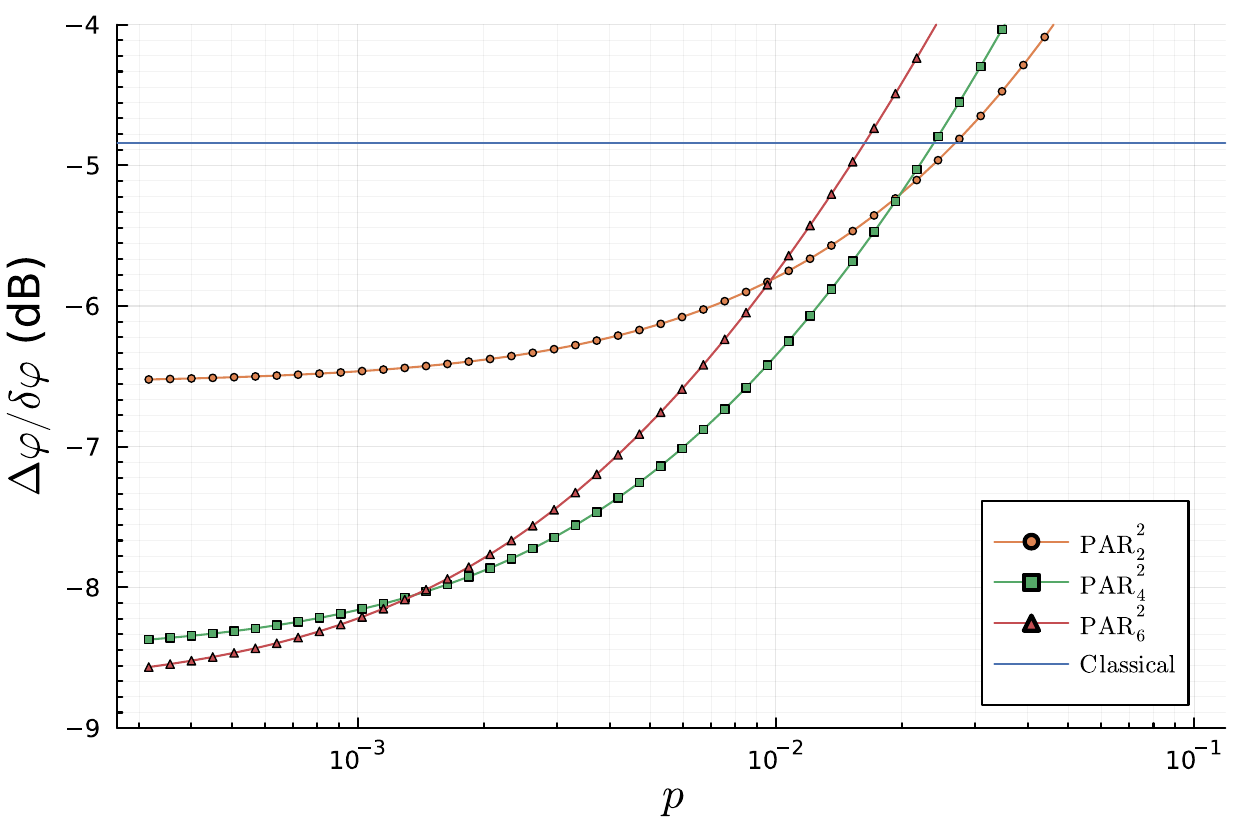}
\put (-1,60) {(c)}
\end{overpic}
\begin{overpic}[width=0.49\textwidth]{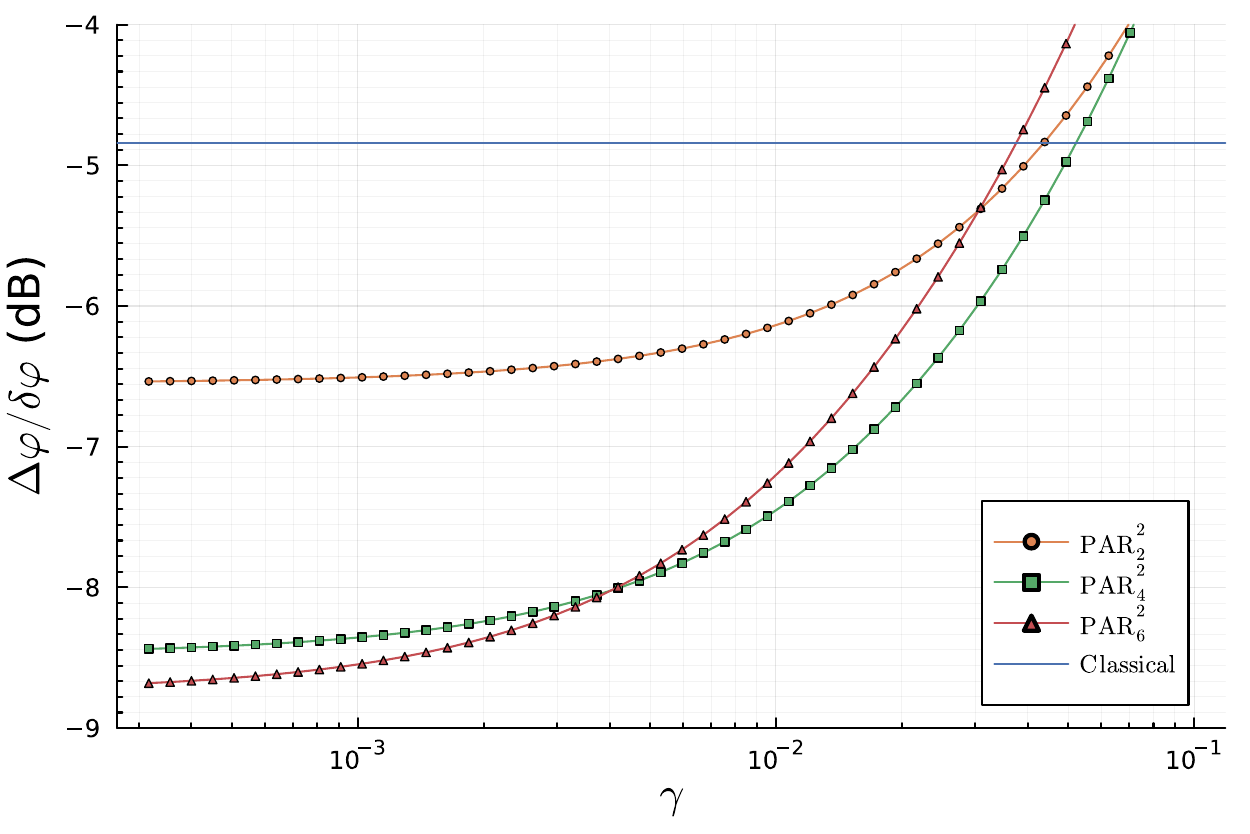}
\put (-1,60) {(d)}
\end{overpic}
\caption{The root Bayesian mean squared error $\Delta\varphi/\delta\varphi$ for ((a) and (b)) $\textrm{AAT}^{1}_{1}$, $\textrm{AAT}^{1}_{2}$, $\textrm{AAT}^{1}_{3}$, $\textrm{AAT}^{1}_{4}$, $\textrm{AAT}^{1}_{5}$, $\textrm{AAT}^{1}_{6}$, ((c) and (d)) $\textrm{PAR}^{2}_{2}$, $\textrm{PAR}^{2}_{4}$, and $\textrm{PAR}^{2}_{6}$, and the classically optimal strategy under noisy one-axis twists vs. the noise strength when $N=30$ and $\delta\varphi\approx 0.74$. In (a) and (c) the noise model is dephasing noise with noise strength $p$ and in (b) and (d) the noise model is amplitude damping noise with noise strength $\gamma$. The ansatzes with deeper decoding circuits are advantageous only when the noise strength $p$ or $\gamma$ is sufficiently small.}
\label{fig:gate}
\end{figure*}

We now study the robustness of the strategies found to be optimal in Sec.~\ref{sec:ansatzes} to various types of noise. First, in the presence of spatially correlated dephasing as described in Sec.~\ref{sec:corr}, we analyze how the performance of the solutions found to be optimal in the absence of noise decays. In particular, we take the correlation functions to be
\begin{equation}
\overline{r_{j}r_{k}}=c_{1}\delta_{j,k}+c_{2}(\delta_{j,k-1}+\delta_{j,k+1}), 
\end{equation}
so that $c_{1}$ is the variance at each site and $c_{2}$ is the covariance between adjacent sites. These results were obtained using our tensor network simulation scheme. Due to the polynomial bond dimensions of the involved states and operators described in Sec.~\ref{sec:tensor} there is no truncation of singular values and the results are exact to numerical precision.

The effect of this type of noise on the optimal $\textrm{AAT}^{1}_{1}$ noiseless strategy found at $\delta\varphi\approx 0.74$ for different values of $c_{1}$ and $c_{2}$ is shown in Fig.~\ref{fig:free}. Note that while we use the circuit parameters from the noiseless optimization, the proportionality constant $a$ in the estimator is changed to be the optimal one given the noise channel defined by Eq.~(\ref{eq:noise}). We do not observe a dramatic drop off in performance in the presence of correlated noise. This is indicated by the relatively constant value of the Bayesian mean squared error along curves of constant $c_{1}+c_{2}$ when $c_{2}>0$. We also find that the Bayesian mean squared error at constant values of $c_{1}$ decreases as $c_{2}$ is made more negative. In other words, performance is improved as the noise becomes anti-correlated. This can be understood as resulting from the underlying estimator being proportional to $\langle J_{z}\rangle$ when averaged over measurement outcomes. Intuitively, as the estimator is the average of a separate estimator for each qubit, we may imagine some cancellation of the errors if the errors in these estimators are anti-correlated.

\subsection{Circuit level noise}

Finally, we study the robustness of the strategies found to be optimal in Sec.~\ref{sec:ansatzes} to circuit level noise, namely noises during encoding and decoding circuits. First, we consider the effect of single-qubit dephasing,
\begin{equation}
\mathcal{D}^{(j)}_{p}[\rho]=(1-p)\rho+pZ_{j}\rho Z_{j},
\end{equation}
on each qubit after each one-axis twist, as a model of a noisy one-axis twist. We refer to $p$ as the noise strength. We again employ the tensor network based simulation and again the proportionality constant $a$ used in the estimator is the optimal one given the noise. The results are shown in Fig.~\ref{fig:gate}a. We observe that $\textrm{AAT}^{1}_{1}$ and $\textrm{AAT}^{1}_{2}$ outperform the classical strategy up to a noise strength of at least $p=0.1$. Second, notice that $\textrm{AAT}^{1}_{2}$ performs better than $\textrm{AAT}^{1}_{1}$ over the same range. This is not a property shared by the ansatzes with deeper decoding circuits. The deeper circuit ansatzes are all outperformed by $\textrm{AAT}^{1}_{2}$ starting at some noise strength between $0.0025$ and $0.0047$. $\textrm{AAT}^{1}_{3}$ is outperformed by $\textrm{AAT}^{1}_{1}$ at $p\approx 0.03$ and by the classical strategy at $p\approx0.079$. The ansatzes with deeper decoding circuits are outperformed by $\textrm{AAT}^{1}_{1}$ beginning between a noise strength of $p\approx0.0085$ and $p\approx0.012$ and they are outperformed by the classical strategy beginning between at noise strength between $p\approx0.035$ and $p\approx0.05$. In Fig.~\ref{fig:gate}c, the robustness of the parity symmetric ansatzes to dephasing noise is displayed. The behavior is qualitatively similar. We note that $\textrm{AAT}^{1}_{1}$ always outperforms $\textrm{PAR}^{2}_{2}$, $\textrm{AAT}^{1}_{4}$ always outperforms $\textrm{PAR}^{2}_{4}$ and $\textrm{PAR}^{2}_{4}$, and $\textrm{AAT}^{1}_{2}$ outperforms $\textrm{PAR}^{2}_{4}$ and $\textrm{PAR}^{2}_{6}$ for noise strengths greater than about $0.00091$. In particular, for noise strengths greater than $0.00091$ all of the considered parity symmetric ansatzes are outperforms by atleast one of $\textrm{AAT}^{1}_{1}$ or $\textrm{AAT}^{1}_{2}$. All of the parity symmetric strategies are outperformed by the classical strategy for error rates greater than $p\approx0.027$ due to the increased number of one-axis twists.

Next, we consider the effect of single qubit amplitude damping. This could arise due to atomic spontaneous emission. The noise channel is given by
\begin{equation}
\begin{split}
\mathcal{A}^{(j)}_{\gamma}[\rho]=&\frac{(1+\sqrt{1-\gamma})^{2}}{4}\rho+\frac{(1-\sqrt{1-\gamma})^{2}}{4}Z_{j}\rho Z_{j} \\
&+\frac{\gamma}{4}\{Z_{j},\rho\}+\gamma\sigma^{(-)}_{j}\rho\sigma^{(+)}_{j},
\end{split}
\end{equation}
where $\gamma$ is the noise strength for this model and $\sigma^{(\pm)}_{j}=\frac{1}{2}(X_{j}\mp iY_{j})$. Again we consider the situation in which each qubit passes through this channel after each one-axis twist. As for the last noise model, the estimator coefficient used is the optimal one given the noise. The results are shown in Fig.~\ref{fig:gate}b. For this noise model $\textrm{AAT}^{1}_{1}$ still outperforms the classically optimal strategy up to at least $\gamma=0.1$. All deeper ansatzes are outperformed by $\textrm{AAT}^{1}_{1}$ at some noise strength between $0.0095$ and $0.028$ and by the classical strategy at some noise strength between $0.044$ and $0.089$. The robustness of the optimized parity symmetric strategies to the same noise is shown in Fig.~\ref{fig:gate}d. Again, the results are qualitatively similar but we note that $\textrm{AAT}^{1}_{1}$ always outperforms $\textrm{PAR}^{2}_{2}$ and $\textrm{AAT}^{1}_{2}$ outperforms $\textrm{PAR}^{2}_{4}$ and $\textrm{PAR}^{2}_{6}$ for noise strengths greater than $0.0047$. Above this noise strength all parity symmetric ansatzes are outperformed by atleast one of $\textrm{AAT}^{1}_{1}$ or $\textrm{AAT}^{1}_{2}$. All of the parity symmetric strategies are outperformed by the entanglement free strategy for values of $\gamma$ greater than $\approx0.0056$.

The first point of this analysis is to give some indication of when it is worth increasing the circuit depth in a particular device as which circuit preforms the best can be dependent on the noise level. The second is to indicate that the presence of noise during the encoding and decoding can rapidly eliminate the benefit of increased decoding depth. This is another sense in which going to deeper ansatzes provides diminishing returns. 

\section{Conclusion}

We have advanced several aspects of Bayesian, or global, variational quantum metrology. We introduced and analyzed a new family of parameterized circuits which we call arbitrary-axis twist ansatzes. These are the most general ansatzes composed of one-axis twists and global rotations with a fixed number of one-axis twists in the encoding and decoding circuits. We found that a reduction in the number of one-axis twists needed for performance comparable to that of a previously considered variational approach based on parity symmetric ansatzes is available. This is especially meaningful in the shallow depth regime where the reduction can be by as much as half. The significance of this finding is complemented by another finding that the usefulness of deeper circuits is harmed by the presence of noise in the encoding and decoding circuits. In the example shown in Fig.~\ref{fig:gate}, we found that for dephasing noise strengths greater than $p=0.00091$ or amplitude damping noise strengths greater than $\gamma=0.0047$, all parity symmetric ansatzes considered are outperformed by $\textrm{AAT}^{1}_{2}$. At larger noise strengths, they are also all outperformed by $\textrm{AAT}^{1}_{1}$. In addition to being the ansatzes utilizing the fewest one-axis twists, $\textrm{AAT}^{1}_{1}$ and $\textrm{AAT}^{1}_{2}$ are usually also among the ansatzes with the smallest total twisting angles (see App.~\ref{sec:time}). In fact, for the example in Fig.~\ref{fig:gate} they are the two ansatzes with the smallest total twisting angle. They also start outperforming the other $\textrm{AAT}$ ansatzes at similar noise strengths. Minimizing the number of entangling resources is also useful from a classical simulation perspective. For example, we simulated the effect of spatially correlated dephasing noise during the free evolution using a tensor network algorithm. The cost of this algorithm was reduced due to the small number of entangling one-axis twist gates in the ansatz of interest. 

Our findings first show the rather important contributions of non-entangling resources to variational metrology especially in the shallow depth regime. This could be a useful fact in the design of future ansatzes. Second, we found that variational metrology can still perform well in the presence of a small amount of spatially correlated noise. Finally, we also studied the robustness of the ideal strategies to circuit level noise, and found that for deep decoding circuits the performance falls off rapidly with the strength of the noise associated with the one-axis twists. This suggests that in the absence of some form of quantum error correction, variational metrology circuits should be kept shallow.

\section*{Acknowledgments}
This work is supported by the National Science Foundation QLCI Q-SEnSE Grant (No. OMA- 2016244), and STAQ Project (No. PHY-1818914). We would like to thank the UNM Center for Advanced Research Computing, supported in part by the National Science Foundation, for providing the high performance computing resources used in this work.

\appendix

\section{Numerical details}\label{sec:numerics}
We numerically search for the optimal values of the circuit parameters for a given value of $\delta\varphi$ for both the arbitrary-axis twist and parity symmetric ansatzes. We first perform the optimization for $\textrm{AAT}^{1}_{1}$ and $\textrm{PAR}^{2}_{2}$. Then the circuit parameters found via this optimization are used as the initialization for one level deeper circuits, e.g. $\textrm{AAT}^{1}_{2}$. The parameters found via this optimization are then used as the starting point for the next level deeper optimization and so on. We refer to this initialization strategy as sequential initialization.

The optimization of each ansatz begins with a Nelder-Mead search~\cite{Nelder65, Box65, Richardson73}. For an optimization over $n$ variables Nelder-Mead constructs an $n+1$ point simplex. Roughly speaking, at each step of the optimization it attempts to replace the point in the simplex with the largest objective value with a better point. If a better point cannot be found, this is interpreted as indicating the presence of a valley and the simplex is shrunk. The optimization is stopped when the difference between the largest and smallest value currently in the simplex differ by less than some value $\epsilon_{1}$.

We then take the output parameters from the Nelder-Mead search and use them as the initialization parameters for a sequential quadratic programming optimization~\cite{Kraft88, Kraft94}. At each iteration, this algorithm finds the minimum of a quadratic approximation to the objective function at the current point. While this approach can provide better performance than linear approximations, it does require access to second derivatives. The stopping criteria is that the minimum objective function value between consecutive iterations differ by less than $\epsilon_{2}$. We then take the output circuit parameters from that optimization and use them to initialize an additional Nelder-Mead optimization. This process is repeated until consecutive optimizations agree to within $\epsilon_{3}$. 

\begin{figure}
\includegraphics[width=0.49\textwidth]{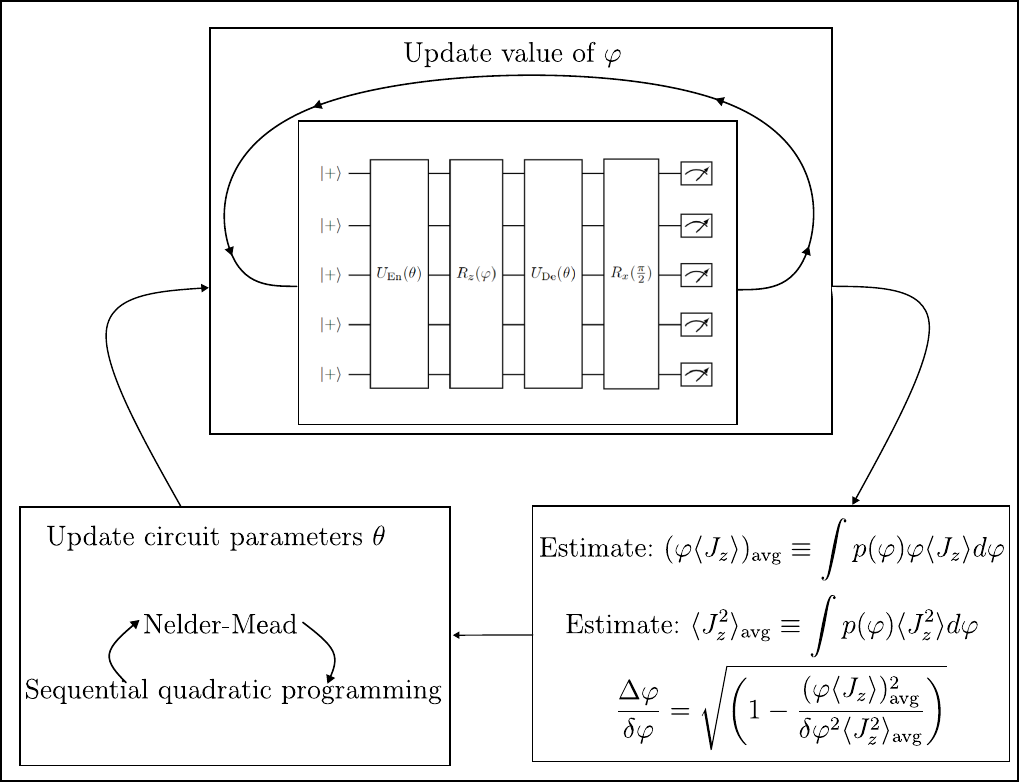}
\caption{Illustration of the algorithm used to find the optimal strategies. First, the expectation values of $J_{z}$ and $J_{z}^{2}$ are estimated at a sufficient number of different values of $\varphi$ to estimate $\Delta\varphi$. Then the estimate of $\Delta\varphi/\delta\varphi$ and possibly estimates of the derivatives of this function are used to choose new values of the circuit parameters $\theta$ to consider.}
\label{fig:flow}
\end{figure}

While the values of $\epsilon_{1}$, $\epsilon_{2}$, and $\epsilon_{3}$ could in principle be allowed to differ, we fix them all to the same value $\epsilon=\epsilon_{1}=\epsilon_{2}=\epsilon_{3}$ and take that value to be $\epsilon=10^{-13}$. The optimization is carried out using the software package NLopt~\cite{nlopt} with automatic differentiation provided by Zygote~\cite{Zygote.jl-2018}. We also note that the numerical integration over $\varphi$ to estimate the Bayesian mean squared error is preformed using Hermite-Gauss quadrature with 500 integration points during the optimization and 25 integration points in the tensor network simulations. The 25 integration points is close to the amount used in a recent proof of principle experiment~\cite{Marciniak_2022}. The total algorithm is illustrated by the flow chart in Fig.~\ref{fig:flow}. The Hermite-Gauss quadrature is preformed using FastGaussQuadrature package~\cite{fgq} and the tensor network simulations use the ITensors package~\cite{itensor, itensor-r0.3}.

\section{Resource counting as measured by the number of one-axis twists vs. total twisting angle}\label{sec:time}

To place in context the method of resource counting utilized in this paper, we briefly comment on the total twisting angle $\theta^{\textrm{twist}}_{\textrm{tot}}$, which is given by the sum of the magnitudes of the angles associated with all one-axis twists used in the ansatz, as a potential alternative to the number of one-axis twists. 
Although the total twisting angle is a seemingly simple way to quantify the cost of the variational strategies for many experimental setups, in this appendix, we will explain why the number of one-axis twists is indeed both a more suitable and more tractable resource to count.
We also note that the strategies considered here do not exhibit large total twisting angles in practice, in fact they are usually less than $\frac{\pi}{8}$. 

\begin{figure}
\includegraphics[width=0.49\textwidth]{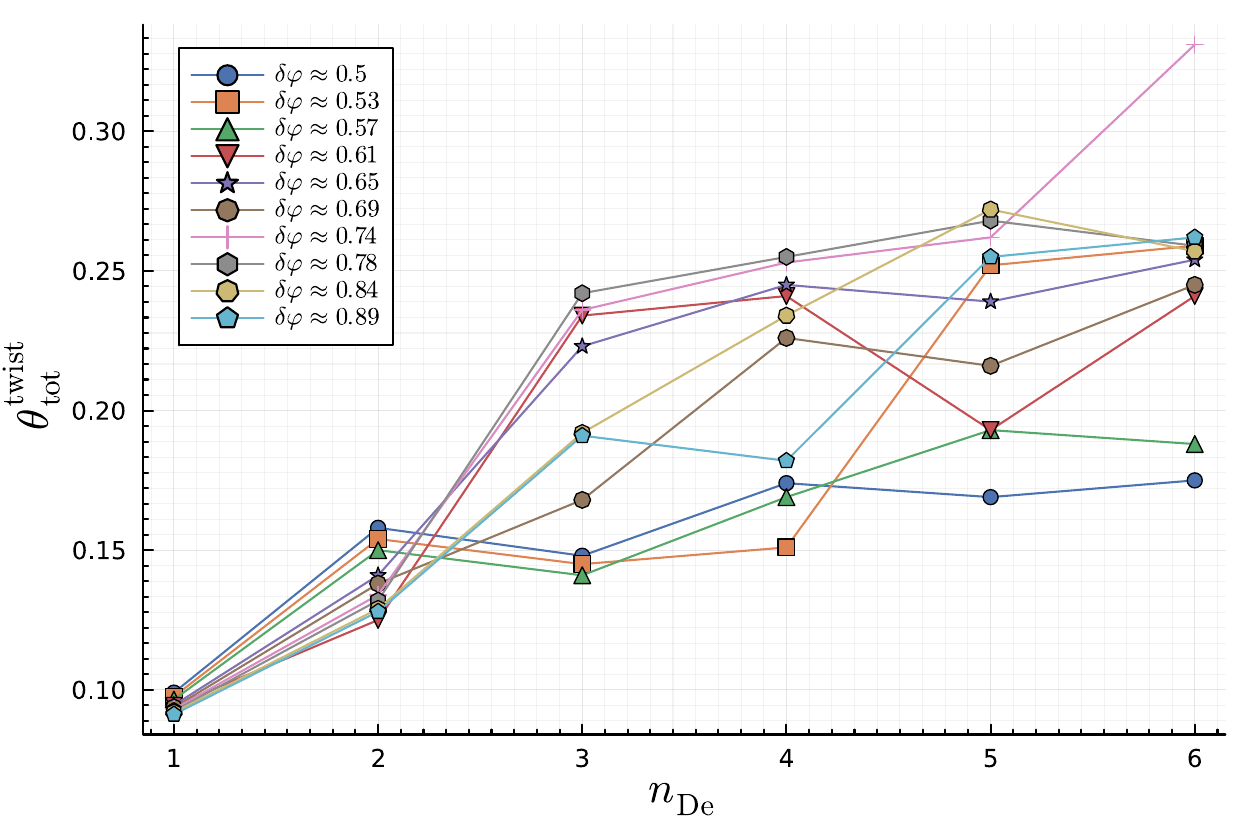}
\caption{Plot of the total twisting angle of the optimal $\textrm{AAT}^{1}_{n_{\textrm{De}}}$ strategies vs. the decoding depth for various values of $\delta\varphi$. Note that there is a clear relation between the two quantities and in the case that $\delta\varphi\approx0.74$ the relationship is nearly linear for the depths considered here.}
\label{fig:time}
\end{figure}

First, in Table.~\ref{tab:twist}, we separately display the total twisting angle associated with the encoding circuit $\theta^{\textrm{twist}}_{\textrm{En}}$ and the total twisting angle associated with the decoding circuit $\theta^{\textrm{twist}}_{\textrm{De}}$ for $N=30$ and $\delta\varphi\approx0.74$ (our most commonly studied prior standard deviation in the main text). The data suggests that these should be treated as two separate objects rather than two parts of $\theta^{\textrm{twist}}_{\textrm{tot}}$. For example, when the data from this table is fit to a curve of the form
\begin{equation}
f(n_{\textrm{En}/\textrm{De}})=An_{\textrm{En}/\textrm{De}}+B,
\end{equation}
$\theta^{\textrm{twist}}_{\textrm{En}}$ exhibits almost constant behavior with $A\approx0.0017$ and $B\approx0.059$ while $\theta^{\textrm{twist}}_{\textrm{De}}$ exhibits a stronger dependence on $n_{\textrm{De}}$ with $A\approx0.044$ and $B\approx0.00041$. This is closely related to the observation in the main text that most of the improvement comes from adding twists to the decoding once a single twist is used in the encoding. In Fig.~\ref{fig:time}, we plot the total twisting angle vs. decoding depth for the optimized $\textrm{AAT}$ ansatzes with $N=30$, $n_{\textrm{En}}=1$, and various values of $\delta\varphi$.  Usually, $\textrm{AAT}$ ansatzes with a larger number twists result in a larger total twisting angle than $\textrm{AAT}$ ansatzes with fewer twists. In the cases where this is not true we often observe little improvement due to the additional twists. These observations suggest that the number of one-axis twists used by the strategy is indeed a reasonable proxy for the total twisting angle for the $\textrm{AAT}$ ansatzes.

Second, the optimization where the number of twists used in the encoding and in the decoding are separately allowed to vary but the total twisting angle is fixed is possible but could lead to unusual solutions if the number of twists is truly unbounded. In particular, such an approach could result in pathological solutions where the number of twists used becomes very large but the angle of each twist becomes very small. Such a strategy may be challenging to implement in some experimental platforms, for example in ion traps \cite{Marciniak_2022}. Additionally, it is still guaranteed that at a fixed total twisting angle the $\textrm{AAT}$ ansatzes will outperform the $\textrm{PAR}$ ansatzes as the later are a special case of the former. We also note that fixing the number of twists does provide an upper bound on the total twisting angle used for that optimization. In this sense, the number of twists can be viewed as a coarse-grained measure to capture much of the physics of the dependence of the BMSE on the magnitude of the twisting angles. For completeness, we mention the total twisting angle for the $\textrm{PAR}$ ansatzes for $\delta\varphi\approx0.74$. The total twisting angles associated with $\textrm{PAR}^{2}_{2}$ and $\textrm{PAR}^{2}_{4}$ are both about $0.153$, and the total angle associated with $\textrm{PAR}^{2}_{6}$ is about $0.194$. These total twisting angles are larger than that of $\textrm{AAT}^{1}_{2}$ but smaller than that of $\textrm{AAT}^{1}_{4}$. However, the total twisting angle varies for other values of $\delta\varphi$. In particular, the total twisting angle of $\textrm{AAT}^{1}_{4}$ has a range between 0.15 and 0.26 as seen in Fig.~\ref{fig:time}, so that it can be comparable to that of $\textrm{PAR}^{2}_{6}$. In particular, for some values of $\delta\varphi$, $\textrm{AAT}^{1}_{4}$ exhibits a smaller total twisting angle than $\textrm{PAR}^{2}_{6}$.

\begin{table}
\centering
\begin{tabular}{c c c c}
\hline\hline
$(n_{\textrm{En}},n_{\textrm{De}})$ & $\theta^{\textrm{twist}}_{\textrm{En}}$ & $\theta^{\textrm{twist}}_{\textrm{De}}$ & $\theta^{\textrm{twist}}_{\textrm{tot}}$ \\
\hline 
(1,1) & 0.0600 & 0.0331 & 0.0931 \\ 
(1,2) & 0.0630 & 0.0713 & 0.134 \\
(1,3) & 0.0642 & 0.172 & 0.236 \\
(1,4) & 0.0675 & 0.186 & 0.253 \\
(1,5) & 0.0680 & 0.194 & 0.262 \\ 
(1,6) & 0.0685 & 0.262 & 0.331 \\
\hline\hline
\end{tabular}
\caption{This table contains the values of $\theta^{\textrm{twist}}_{\textrm{En}}$, $\theta^{\textrm{twist}}_{\textrm{De}}$, and $\theta^{\textrm{twist}}_{\textrm{tot}}$ for $n_{En}=1$, $N=30$, $\delta\varphi\approx0.74$, and various values of $n_{\textrm{De}}$. The values are reported here to three significant figures.}
\label{tab:twist}
\end{table}

\section{Comparison of arbitrary-axis twist and parity symmetric ansatzes}\label{sec:compare}

\begin{figure*}
\begin{overpic}[width=0.49\textwidth]{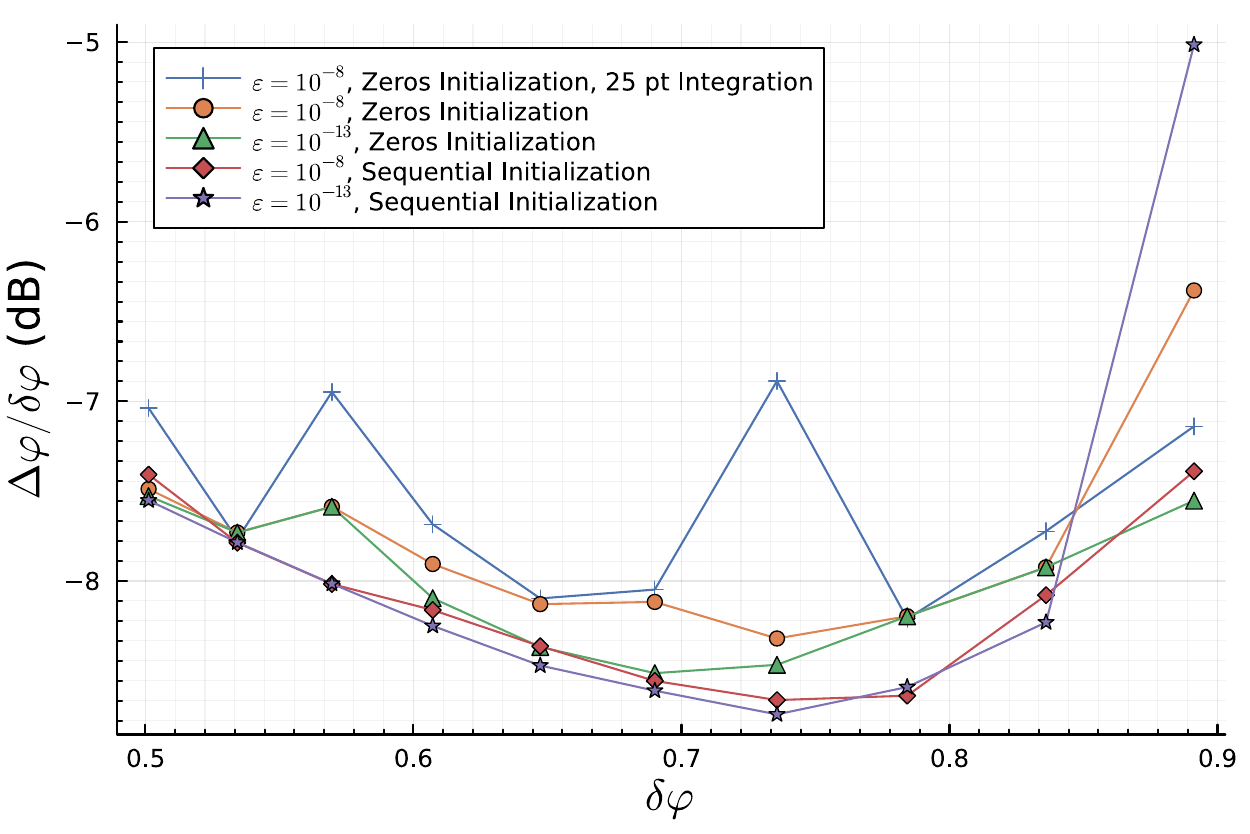}
\put (-1,60) {(a)}
\end{overpic}
\begin{overpic}[width=0.49\textwidth]{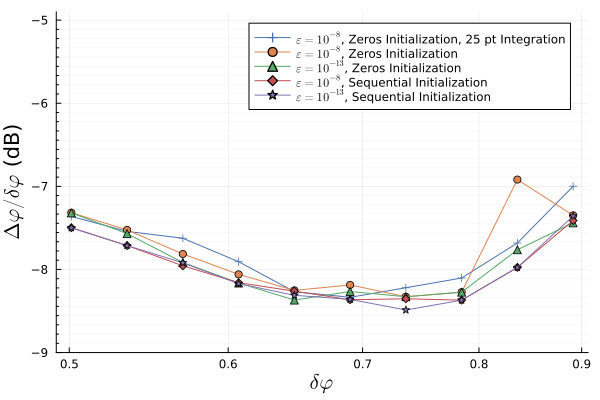}
\put (-1,60) {(b)}
\end{overpic}
\caption{Plot of the root Bayesian mean squared error $\Delta\varphi/\delta\varphi$ vs. prior variance $\delta\varphi$ resulting from the Nelder-Mead/sequential quadratic programming approach described in the text for various hyperparameters for (a) $\textrm{PAR}^{2}_{6}$ and (b) $\textrm{AAT}^{1}_{3}$. The blue crosses are the results for initializing all circuit parameters to zero, setting $\epsilon=10^{-8}$, and using only 25 points in the numerical integration. The orange circles are the same as the blue crosses but with 500 points used in the numerical integration. The green triangles are the same as the orange circles but with $\epsilon=10^{-13}$. The red diamonds use $\epsilon=10^{-8}$ but sequential initialization. Finally, the purple triangles are the result of using the hyperparameters described in App.~\ref{sec:numerics}. As they are usually the best, the numerical setting of the purple triangles are displayed throughout the main text.}
\label{fig:opt_meth}
\end{figure*}

We note that the result of the optimization, and even which ansatzes outperform others, can depend significantly on the specifics of the hyperparameters used. To study this phenomena we perform the optimization of $\textrm{PAR}^{2}_{6}$ and $\textrm{AAT}^{1}_{3}$ for different sets of hyperparameters. In particular, the hyperparameters we adjust are $\epsilon$, whether the sequential initialization is used or if all circuit parameters are initialized to zero, and the number of points used to approximate the integral over $\varphi$. The results of all these optimizations are shown in Fig.~\ref{fig:opt_meth}. We chose these two ansatzes as examples for three reasons. First, they are fairly deep circuits. Second, $\textrm{PAR}^{2}_{6}$ and $\textrm{AAT}^{1}_{3}$ have a similar number of circuit parameters with 12 and 16 respectively. Third, they are an example where which ansatz achieves a smaller BMSE depends on the hyperparameters used. We find that the optimizations are liable to get stuck in local minimia whenever the hyperparameters are relaxed from the values described above. 

This sensitivity to hyperparameters has the potential to be an issue for experimental on device optimization since it necessitates examining more values of $\varphi$ for the integration and using more repetitions to increase the estimation precision of $\Delta\varphi/\delta\varphi$. We also note that we have found that the sequential initialization strategy described above is able to at least partially mitigate these issues. We have also heuristically observed that the arbitrary-axis twist ansatzes may be somewhat less susceptible to these issues than the parity symmetric ansatzes. On an optimistic note, most optimizations found reasonably good values for $\Delta\varphi/\delta\varphi$ even if they do not find the global minimum. This likely acceptable for some applications.

\section{MPOs for permutation invariant operators}\label{sec:mpo}
In this appendix, we provide a construction of an MPO for any permutation invariant operator with bond dimension $O(N^{3})$. The action of the permutation group defines a set of equivalence classes of computational basis states labeled by the Hamming weight according to
\begin{equation}
|x\rangle\sim|y\rangle\quad\textrm{iff}\ \exists\sigma\in S_{N}\ \textrm{s.t.}\ x=\sigma(y).
\end{equation}
For reasons that will become clear shortly, in this appendix and the next we will denote the Hamming weight as $t_{1}$. It is useful to choose a representative element from each of these equivalence classes. We choose elements of the form $|x(\mathbf{t})\rangle=|0\rangle^{\otimes N-t_{1}}\otimes |1\rangle^{\otimes t_{1}}$ so that the first $N-t_{1}$ qubits are in state $|0\rangle$ and the rest are in state $|1\rangle$. A permutation invariant state $|\psi\rangle$ is then one of the form
\begin{equation}
|\psi\rangle=\sum_{\substack{t_{1}\\ \textrm{s.t.}\ t_{1}\leq N}}\frac{c_{\mathbf{t}}}{t_{1}!(N-t_{1})!}\sum_{\sigma\in S_{N}}P_{\sigma}|x(\mathbf{t})\rangle,
\end{equation}
where the denominator of the coefficient is a normalization convention which accounts for repetitions of identical kets and $P_{\sigma}|x\rangle\equiv|\sigma(x)\rangle$. 

This framework can be naturally extended to states of $N$ qudits. We denote a standard basis for qudits of dimension $d$ by $\{|0\rangle,|1\rangle,\dots,|d-1\rangle\}$. A standard basis state for $N$ qudits of dimension $d$ then has the form $\otimes_{j}|s_{j}\rangle\equiv|s_{1}s_{2}\cdots s_{N}\rangle$ where each $|s_{j}\rangle\in\{|0\rangle,|1\rangle,\dots,|d-1\rangle\}$. Now the equivalence classes defined by the action of the permutation group are labeled by a type vector of the form $\mathbf{t}=(t_{1},t_{2},\dots,t_{d-1})$ subject to the constraint $\sum_{j=1}^{d-1}t_{j}\leq N$. We again choose a representative element from each equivalence class as $|x(\mathbf{t})\rangle=|0\rangle^{\otimes N-\sum_{j}t_{j}}\otimes|1\rangle^{\otimes t_{1}}\otimes\cdots\otimes|d-1\rangle^{\otimes t_{d-1}}$. It is convenient to define $t_{0}\equiv N-\sum_{j}t_{j}$ to be the number of qudits in state $|0\rangle$. Then permutation invariant qudit states $|\psi\rangle$ can be written as
\begin{equation}
|\psi\rangle=\sum_{\substack{\mathbf{t}\\ \textrm{s.t.}\ \sum_{j}t_{j}\leq N}}\frac{c_{\mathbf{t}}}{t_{0}!\prod_{j}t_{j}!}\sum_{\sigma\in S_{N}}P_{\sigma}|x(\mathbf{t})\rangle.
\end{equation}
The number of independent coefficients here is equal to the number of distinct type vectors, or the number of multisets of cardinality $N$ with elements drawn from an underlying set of cardinality $d$, i.e. ${N+d-1\choose d-1}=O(N^{d-1})$.

This framework can also include permutation invariant operators via the Choi-Jamiołkowski isomorphism. Under this isomorphism, an operator $\mathcal{O}=\sum_{s,s'}c_{s,s'}|s\rangle\langle s'|$ on qudits of dimension $d$ is associated with a state of qudits of dimension $d^{2}$ according to
\begin{equation}
\begin{split}
\mathcal{O}&=\sum_{s,s'}c_{s,s'}|s\rangle\langle s'| \\
&\rightarrow|\mathcal{O}\rangle=\sum_{s,s'}c_{s,s'}|(s_{1},s_{1}')(s_{2},s_{2}')\cdots(s_{N},s_{N}')\rangle,
\end{split}
\end{equation}
where the standard $d^{2}$ dimensional qudit states are now labeled by the ordered pairs $(s_{j},s_{j}')$ and the $s_{j}$ can take $d$ values. In order make the consequences for the underlying operator transparent, when considering an operator on qubits we will write the type vectors labeling the equivalence classes as $\mathbf{t}=(t_{(0,1)},t_{(1,0)},t_{(1,1)})$ and identify a representative from each equivalence class as $|x(\mathbf{t})\rangle=|(0,0)\rangle^{\otimes N-t_{(0,1)}-t_{(1,0)}-t_{(1,1)}}\otimes|(0,1)\rangle^{\otimes t_{(0,1)}}\otimes|(1,0)\rangle^{\otimes t_{(1,0)}}|(1,1)\rangle^{\otimes t_{(1,1)}}$. This representative element can be connected via the Choi-Jamiołkowski isomorphism to an operator on qubits,
\begin{equation}
\begin{split}
\hat{x}(\mathbf{t})=|0\rangle\langle 0|&^{\otimes N-t_{(0,1)}-t_{(1,0)}-t_{(1,1)}}\\
&\otimes|0\rangle\langle 1|^{\otimes t_{(0,1)}}\otimes|1\rangle\langle 0|^{\otimes t_{(1,0)}}\otimes|1\rangle\langle 1|^{\otimes t_{(1,1)}}.
\end{split}
\end{equation}
A permutation invariant operator is then one that can be written as
\begin{equation}
\mathcal{O}=\sum_{\substack{\mathbf{t} \\ \textrm{s.t.}\ \sum_{j}t_{j}\leq N}}\frac{c_{\mathbf{t}}}{t_{(0,0)}!\prod_{j}t_{j}!}\sum_{\sigma\in S_{N}}P_{\sigma}\hat{x}(\mathbf{t})P^{\dag}_{\sigma},
\end{equation}
where, in analogy with the qudit case, $t_{(0,0)}\equiv N-\sum_{j}t_{j}$.

It has been known for a long time that permutation invariant states of qubits can be represented with bond dimension that scales linearly with system size~\cite{Perez-Garcia_2007}. Here we give a concrete construction that works for any state of qudits of any dimension $d$ in the symmetric subspace and has maximum bond dimension $O(N^{d-1})$. That is, the resulting MPS has bond dimension that is polynomial in the system size. This construction can be extended to MPOs for operators via the Choi-Jamiołkowski isomorphism. 

In this construction, the tensor at a site $1$ is described by $1\times d$ matrices of the form, $(A^{(s)}[1])_{1,m}=\delta_{s,m-1}$. This vector can be thought of as keeping track of the type vector of some given basis ket in the sense that there is one possible vector with a single entry equal to one and the rest zero here for each possible type vector of a one site system. The remaining tensors associated with sites on the left half of the chain, i.e. those sites with $j<\lceil\frac{N+1}{2}\rceil$ will serve to update this vector with new information about the global type as we move from left to right in that $\prod_{k=1}^{j}A^{(s_{k})}[k]$ should be a $1\times {j+d-1 \choose d-1}$ vector in which each entry is associated with one of the possible type vectors of a $j$ site system. Accordingly, the right dimension of $A^{(s)}[j]$ is ${j+d-1 \choose d-1}=O(j^{d-1})$. The sorting of the entries of this vector is in order of increasing values of $||\mathbf{t}||_{1}=\sum_{j}t_{j}$. For types with the same value of $||\mathbf{t}||_{1}$, the entries are sorted first in order of decreasing $t_{1}$. For types with the same value of $t_{1}$ they are sorted in order of decreasing $t_{2}$  and so on. 

If $s=0$, then $(A^{(0)}[j])_{\mu,\nu}=\delta_{\mu,\nu}$. When $s\neq0$, the tensors are constructed recursively as $(A^{(s)}[j])=[A^{(s)}[j-1], C_{j}^{s}]$ where the matrix $C_{j}^{s}$ has right dimension ${j+d-2 \choose d-2}$ and all of its elements are either zero or one. The non-zero elements of the matrix $(C^{s}_{j})_{\mu,\nu}$ are those whose components satisfy
\begin{align}
\begin{split}
\mu&=\sum_{k=1}^{d-2}\sum_{n_{k}=0}^{(j-1)-\sum_{\ell=1}^{k}t_{\ell}-1} {n_{k}+d-2-k \choose d-2-k}+1,
\end{split}
\\[2ex]
\begin{split}
\nu&=\sum^{s-1}_{k=1}\sum_{n_{k}=0}^{j-\sum_{\ell=1}^{k}t_{\ell}-1}{n_{k}+d-2-k \choose d-2-k} \\
&\quad +\sum^{d-2}_{k=s}\sum_{n_{k}=0}^{j-\sum_{\ell=1}^{k}t_{\ell}-2}{n_{k}+d-2-k \choose d-2-k}+1
\end{split}
\end{align}
for some choice of $\{t_{n}\}$ satisfying $\sum_{n}t_{n}=j-1$. The tensors for the right half of the chain $j>\lceil\frac{N+1}{2}\rceil$ are constructed in the same way but with the role of column and row indices exchanged, i.e. they are the transpose.

This leaves a single site unaccounted for at $j=\lceil\frac{N+1}{2}\rceil$. All the coefficients that are needed to describe this operator can be placed in the tensor at this site. The $\mu\nu$ matrix element of $A^{(s)}[\lceil\frac{N+1}{2}\rceil]$ has the value $c_{\mathbf{t}}$ if
\begin{align}
\begin{split}
\mu&={||\mathbf{t}^{(L)}||_{1}-2+d \choose d-1} \\
&\quad+\sum_{k=1}^{d-2}\sum_{n_{k}=0}^{\sum_{\ell=k+1}^{d-1}t^{(L)}_{\ell}-1}{n_{k}+d-2-k \choose d-2-k}+1,
\end{split}
\\[2ex]
\begin{split}
\nu&={||\mathbf{t}^{(R)}||_{1}-2+d \choose d-1}\\
&\quad+\sum_{k=1}^{d-2}\sum_{n_{k}=0}^{\sum_{\ell=k+1}^{d-1}t^{(R)}_{\ell}-1}{n_{k}+d-2-k \choose d-2-k}+1
\end{split}
\end{align}
with $\mathbf{t}^{(L)}+\mathbf{t}^{(R)}+\hat{e}_{s}=\mathbf{t}$ where $\hat{e}_{s}$ is the type vector with all elements equal to zero except for element $s$ which is equal to one and we use the convention that ${n \choose m}=0$ if $m<0$. The result of this construction is an exact MPS or MPO representation of the state or operator with bond dimension ${\lceil\frac{N+1}{2}\rceil+d-1 \choose d-1}=O(N^{d-1})$ for pure states and ${\lceil\frac{N+1}{2}\rceil +d^{2}-1 \choose d^{2}-1}=O(N^{d^2-1})$ for operators.

\section{Permutation invariant compilation}\label{sec:compile}
Here, we show that a polynomial depth circuit of permutation invariant unitaries can be compiled into a single permutation invariant unitary at a cost polynomial in system size. In App.~\ref{sec:mpo}, we saw that a permutation invariant operator is specified by only ${N+3\choose N}=O(N^{3})$ coefficients, one for each possible length three type vector for $N$ sites. Thus, given two permutation invariant operators $U$ and $V$ with coefficients $a_{\mathbf{t}}$ and $b_{\mathbf{t}'}$ respectively, our task is to find the coefficients $c_{\mathbf{t}''}$ of their product $UV$. We will prime the elements of the type vectors in the same way as the type vectors themselves so that, for example, $\mathbf{t}''=(t''_{(0,1)}, t''_{(1,0)}, t''_{(1,1)})$.

Ultimately, the resulting bra-ket pair on qubit $j$ can end up with the form $|\mu\rangle\langle\nu|$ in only two ways. Either it emerged from a product of the form $|\mu\rangle\langle0|0\rangle\langle\nu|$ or from a product of the form $|\mu\rangle\langle1|1\rangle\langle\nu|$. It will be convenient to label the number of $(\mu,\nu)$ pairs that emerge from the first type as $n^{\mu0}_{0\nu}$. Further, the contribution will be nonzero only if the resulting number of $t''_{(1,0)}$ pairs is equal to the number that emerge from the $|1\rangle\langle0|0\rangle\langle0|$ products plus the number that emerge from the $|1\rangle\langle1|1\rangle\langle0|$, that is $t''_{(\mu,\nu)}=n^{\mu0}_{0\nu}+n^{\mu1}_{1\nu}$. We must also have $t_{(\mu,\nu)}=n^{\mu\nu}_{\nu0}+n^{\mu\nu}_{\nu1}$ and $t'_{(\mu,\nu)}=n^{0\mu}_{\mu\nu}+n^{1\mu}_{\mu\nu}$. This allows us to derive conditions on the type vector of the two input operators in terms of these parameters and the type of the output operator. For the first operator, we have
\begin{align}
t_{(0,1)}&=n^{01}_{10}+n^{01}_{11}=n^{01}_{10}+t''_{(0,1)}-n^{00}_{01}, \\
t_{(1,0)}&=n^{10}_{00}+n^{10}_{01}=t''_{(1,0)}-n^{11}_{10}+n^{10}_{01}, \\
t_{(1,1)}&=n^{11}_{10}+n^{11}_{11}=n^{11}_{10}+t''_{(1,1)}-n^{10}_{01}.
\end{align}
The conditions on the type of the second operator are
\begin{align}
t'_{(0,1)}&=n^{00}_{01}+n^{10}_{01}, \\
t'_{(1,0)}&=n^{01}_{10}+n^{11}_{10}, \\
t'_{(1,1)}&=n^{01}_{11}+n^{11}_{11}=t''_{(0,1)}-n^{00}_{01}+t''_{(1,1)}-n^{10}_{01}.
\end{align}
Thus, the types of both input operators are fixed in terms of the type of the output operator and the four parameters $\{n^{00}_{01},n^{10}_{01},n^{01}_{10},n^{11}_{10}\}$.

It is natural to first consider the product of the sum all possible permutations of two of our representative elements of the equivalence classes, i.e.
\begin{equation}
\begin{split}
\left(\frac{1}{t_{(0,0)}!\prod_{j}t_{j}!}\right)&\left(\frac{1}{t'_{(0,0)}!\prod_{j}t'_{j}!}\right) \\
\times&\sum_{\sigma,\sigma'}P_{\sigma}\hat{x}(\mathbf{t})P^{\dag}_{\sigma}P_{\sigma'}\hat{x}(\mathbf{t}')P^{\dag}_{\sigma'}.
\end{split}
\end{equation}
It is useful to note that the second line of this expression can be rewritten as
\begin{equation}
\begin{split}
\sum_{\sigma\in S_{N}}&P_{\sigma}\left(\hat{x}(\mathbf{t})\sum_{\sigma'\in S_{N}}P_{\sigma^{-1}\sigma'}x(\mathbf{t}')P^{\dag}_{\sigma^{-1}\sigma'}\right)P_{\sigma}^{\dag} \\
&=\sum_{\sigma\in S_{N}}P_{\sigma}\left(\hat{x}(\mathbf{t})\sum_{\tilde{\sigma}\in S_{N}}P_{\tilde{\sigma}}x(\mathbf{t}')P^{\dag}_{\tilde{\sigma}}\right)P_{\sigma}^{\dag}.
\end{split}
\end{equation}
Each term in the sum in the parenthesis will be non-zero only if all of the zero-bras in $\hat{x}(\mathbf{t})$ line up with a zero-ket in $P_{\tilde{\sigma}}x(\mathbf{t}')P^{\dag}_{\tilde{\sigma}}$ and similarly for the one-bras and one-kets. For a fixed values of the parameters $n^{\mu\nu}_{\nu\rho}$. The number of non-zero contributions are
\begin{equation}
\begin{split}
{t_{(0,0)}\choose n^{00}_{01}}&{t_{(0,1)}\choose n^{01}_{10}}{t_{(1,0)}\choose n^{10}_{01}}{t_{(1,1)}\choose n^{11}_{10}} \\
&=\frac{t_{(0,0)}!t_{(0,1)}!t_{(1,0)}!t_{(1,1)}!}{(n^{00}_{00}!n^{01}_{10}!)(n^{00}_{01}!n^{01}_{11}!)(n^{10}_{00}!n^{11}_{10}!)(n^{10}_{01}!n^{11}_{11}!)}.
\end{split}
\end{equation}
The numerator exactly cancels the normalization associated with the type vector $\mathbf{t}$. To obtain the coefficients $c_{\mathbf{t}''}$ we will need to multiply this by the normalization associated with the type vector $\mathbf{t}''$. The resulting factor is
\begin{equation}
\begin{split}
\frac{t''_{(0,0)}!t''_{(0,1)}!t''_{(1,0)}!t''_{(1,1)}!}{t_{(0,0)}!t_{(0,1)}!t_{(1,0)}!t_{(1,1)}!}&{t_{(0,0)}\choose n^{00}_{01}}{t_{(0,1)}\choose n^{01}_{10}}{t_{(1,0)}\choose n^{10}_{01}}{t_{(1,1)}\choose n^{11}_{10}} \\
=&{{t_{(0,0)}''}\choose{n_{10}^{01}}}{{t_{(1,1)}''}\choose{n_{01}^{10}}}{{t_{(0,1)}''}\choose{n_{01}^{00}}}{{t_{(1,0)}''}\choose{n_{10}^{11}}}.
\end{split}
\end{equation}

The coefficients associated with $UV$ are then
\begin{widetext}
\begin{equation}
\begin{split}
c_{\mathbf{t}''}=\sum_{n_{10}^{01}=0}^{t_{(0,0)}''}\sum_{n_{01}^{10}=0}^{t_{(1,1)}''}\sum_{n_{01}^{00}=0}^{t_{(0,1)}''}\sum_{n_{10}^{11}=0}^{t_{(1,0)}''}
{{t_{(0,0)}''}\choose{n_{10}^{01}}}{{t_{(1,1)}''}\choose{n_{01}^{10}}}{{t_{(0,1)}''}\choose{n_{01}^{00}}}{{t_{(1,0)}''}\choose{n_{10}^{11}}} 
&a_{(n_{10}^{01}+t''_{(0,1)}-n_{01}^{00},t''_{(1,0)}-n^{11}_{10}+n_{01}^{10},n_{10}^{11}+t''_{(1,1)}-n_{01}^{10})} \\
&\times b_{(n_{01}^{00}+n_{01}^{10},n^{01}_{10}+n^{11}_{10},t''_{(0,1)}-n^{00}_{01}+t''_{(1,1)}-n^{10}_{01})}.
\end{split}
\end{equation}
\end{widetext}
These sums each run over $N+1$ entries, so the complexity of computing this coefficient is $O(N^{4})$. Since the output operator is specified by $O(N^{3})$ coefficients, the complexity of the multiplication is $O(N^{7})$ in time and $O(N^{3})$ in memory. We emphasize that these coefficients contain all the information about how the operator acts on the entire state space, not just the symmetric subspace.

\bibliography{references}

\end{document}